\documentclass[journal]{IEEEtran}

\usepackage{times, graphicx,amsmath,amsthm,amssymb}
\usepackage{cite}
\usepackage{url}

\usepackage[tight,footnotesize]{subfigure}
\usepackage{bm,nicefrac}
\usepackage{algorithmic}
\usepackage{algorithm,units}

\newcommand{\n}{n}
\newcommand{\inDim}{n}
\newcommand{\outDim}{m}

\newcommand{\Flow}{x}
\newcommand{\hFlow}{\widehat{x}}
\newcommand{\flow}{x}

\newcommand{\ts}{\theta^*}
\newcommand{\that}{\hat{{\theta}}}

\newcommand{\tb}{{\theta}}
\newcommand{\ybs}{{{\beta^*}}}
\newcommand{\ybh}{{{\hat{\beta}}}}
\newcommand{\yb}{{y}}
\newcommand{\Rate}{\lambda}
\newcommand{\Rates}{\Rate^*}
\newcommand{\hRate}{\widehat{\rate}}
\newcommand{\has}{\hRate}
\newcommand{\rate}{\lambda}
\newcommand{\rates}{\rate^*}
\newcommand{\hrate}{\widehat{\rate}}
\newcommand{\Cts}{\yb}
\newcommand{\bound}{D\left\|\rates-{\rates}^{(k)} \right\|_{\infty}}
\def\pen{{\rm pen}}
\def\half{1/2}
\def\dir{{\rm dir}}
\def\pMLE{{\rm pMLE}}
\def\Poi{{\rm Poisson}}
\def\Risk#1#2{R\left( #1, #2\right)} 
\def\kerr#1{\sigma_k(#1)} 

\def\cN{{\mathcal N}}

\def\Graph{G}

\def\Edges{E}
\def\Vin{A} 
\def\Vout{B} 
\def\AdjMat{{F}} 
\def\nAdjMat{{\Phi}} 
\def\eps{\epsilon}
\def\Prob{{\mathbb P}}
\def\Exp{{\mathbb E}}
\def\res{\tau} 
\def\samp{\nu} 
\newcommand{\T}{\samp}

\newcommand{\post}{\textit{a posteriori }}

\newcommand{\m}{{ m}}
\newcommand{\Phii}{{\Phi}}
\newcommand{\Ex}{\mathbb{E}}

\newcommand{\A}{\Phii}
\newcommand{\At}{\tilde{F}}
\newcommand{\RE}{\mbox{KL}}
\newcommand{\reals}{\mathbb{R}}
\newcommand{\R}{\reals}
\newcommand{\Rplus}{\reals_+}
\newcommand{\Zplus}{\mathbb{Z}_+}

\newcommand{\la}{{g}}
\newcommand{\lb}{{h}}

\def\deq{\stackrel{\scriptscriptstyle\triangle}{=}}
\newcommand\argmin{\operatornamewithlimits{argmin}}
\theoremstyle{definition}
\newtheorem{theorem}{Theorem}[section]
\newtheorem{definition}[theorem]{Definition}
\newtheorem{proposition}[theorem]{Proposition}

\newtheorem{remark}[theorem]{Remark}
\newtheorem{lemma}[theorem]{Lemma}

\begin{document}


\title{Performance Bounds for Expander-Based Compressed Sensing in Poisson Noise}

\author{Maxim Raginsky,~\IEEEmembership{Member,~IEEE}, Sina Jafarpour, Zachary~T.~Harmany,~\IEEEmembership{Student Member,~IEEE}, Roummel~F.~Marcia,~\IEEEmembership{Member,~IEEE}, Rebecca~M.~Willett,~\IEEEmembership{Member,~IEEE}, and Robert Calderbank,~\IEEEmembership{Fellow,~IEEE}
\thanks{The work of M.~Raginsky, Z.T.~Harmany, and R.M.~Willett was supported by NSF CAREER Award No.~CCF-06-43947, DARPA Grant No.~HR0011-07-1-003, and NSF Grant DMS-08-11062. The work of R.~Calderbank and S.~Jafarpour was supported in part by NSF under grant DMS 0701226, by ONR under grant
N00173-06-1-G006, and by AFOSR under grant FA9550-05-1-0443. The work
of R.F.~Marcia was supported by NSF Grant No.~DMS-08-11062.}
\thanks{M.~Raginsky is with the Department of Electrical and Computer Engineering, Duke University, Durham, NC 27708 USA (e-mail: m.raginsky@duke.edu).}
\thanks{S.~Jafarpour is with the Department of Computer Science,
    Princeton University, Princeton, NJ 08540 USA (e-mail: sina@cs.princeton.edu).}
\thanks{Z.T.~Harmany and R.M.~Willett are with the
    Department of Electrical and Computer Engineering, Duke
    University, Durham, NC, 27708, USA (e-mail: zth@duke.edu, willett@duke.edu).}
\thanks{R. Marcia is with the School
    of Natural Sciences, University of California, Merced, CA 95343 USA (e-mail: rmarcia@ucmerced.edu).}
\thanks{R.~Calderbank is with the Department of Computer Science, Duke University, Durham, NC 27708 USA (e-mail: robert.calderbank@duke.edu).}}
\maketitle

\begin{abstract}
  This paper provides performance bounds for compressed sensing in the
  presence of Poisson noise using expander graphs. The Poisson noise
  model is appropriate for a variety of applications, including
  low-light imaging and digital streaming, where the
  signal-independent and/or bounded noise models used in the
  compressed sensing literature are no longer applicable. In this
  paper, we develop a novel sensing paradigm based on expander graphs
  and propose a MAP algorithm for recovering sparse or compressible
  signals from Poisson observations. The geometry of the expander
  graphs and the positivity of the corresponding sensing matrices play
  a crucial role in establishing the bounds on the signal
  reconstruction error of the proposed algorithm. We support our results with
  experimental demonstrations of reconstructing average packet arrival
  rates and instantaneous packet counts at a router in a communication
  network, where the arrivals of packets in each flow follow a Poisson
  process.
\end{abstract}


\maketitle

\thispagestyle{empty}

\section{Introduction} 

The goal of \textit{compressive sampling} or \textit{compressed
  sensing} (CS) \cite{donoho,CRT} is to replace conventional sampling
by a more efficient data acquisition framework, which generally
  requires fewer sensing resources. This paradigm is particularly
  enticing whenever the measurement process is costly or constrained in some
  sense. For example, in the context of photon-limited applications
  (such as low-light imaging), the photomultiplier tubes used within
  sensor arrays are physically large and expensive. Similarly, when
  measuring network traffic flows, the high-speed memory used in
  packet counters is cost-prohibitive. These problems appear ripe for
  the application of CS.

However, photon-limited measurements \cite{SnyderCCD} and
  arrivals/departures of packets at a router \cite{BertsekasNetworks}
  are commonly modeled with a Poisson probability distribution, posing
  significant theoretical and practical challenges in the context of
CS. One of the key challenges is the fact that the measurement error
variance scales with the true intensity of each measurement, so that
we cannot assume constant noise variance across the collection of
measurements. Furthermore, the measurements, the underlying true
  intensities, and the system models are all subject to certain physical
  constraints, which play a significant role in performance.

Recent works \cite{expFamCS,dequantizing,impulsive,justice}
  explore methods for CS reconstruction in the presence of impulsive,
  sparse or exponential-family noise, but do not account for the
  physical constraints associated with a typical Poisson setup and
  do not contain the related performance bounds emphasized in this paper. In previous work \cite{wr,rhmw}, we showed that a Poisson noise model
combined with conventional dense CS sensing matrices (properly scaled)
yielded performance bounds that were somewhat sobering relative to
bounds typically found in the literature. In particular, we found that
if the number of photons (or packets) available to sense were held
constant, and if the number of measurements, $m$, was above some
critical threshold, then larger $m$ in general led to larger bounds on
the error between the true and the estimated signals.  This can
intuitively be understood as resulting from the fact that dense CS
measurements in the Poisson case cannot be zero-mean, and the DC
offset used to ensure physical feasibility adversely impacts the noise
variance.

The approach considered in this paper hinges, like most CS methods, on
reconstructing a signal from compressive measurements by optimizing a
sparsity-regularized goodness-of-fit objective function. In contrast
to many CS approaches, however, we measure the fit of an estimate to
the data using the Poisson log-likelihood instead of a squared error
term. This paper demonstrates that the bounds developed in previous
work can be improved for some sparsity models by considering
alternatives to dense sensing matrices with random entries. In
particular, we show that {\em deterministic} sensing matrices given by
scaled adjacency matrices of expander graphs have important
theoretical characteristics (especially an $\ell_1$ version of the
{\em restricted isometry property} \cite{rip}) that are ideally suited to
controlling the performance of Poisson CS.

Formally, suppose we have a signal $\ts \in \reals_+^n$ with known
$\ell_1$ norm $\|\ts\|_1$ (or a known upper
bound on $\|\ts\|_1$). We aim to find a matrix $\Phii \in \reals_+^{m
  \times n}$ with $m$, the number of measurements, as small as
possible, so that $\ts$ can be recovered efficiently from the measured
vector $\yb \in \reals_+^m$, which is related to 
$\Phii \ts$ through a Poisson observation model. The restriction that elements of $\Phii$ be nonnegative
reflects the physical limitations of many sensing systems of interest
(e.g., packet routers and counters or linear optical systems). The
original approach employed dense random matrices \cite{rip,CRT2}. It
has been shown that if the matrix $\Phii$ acts nearly isometrically on
the set of all $k$-sparse signals, thus obeying what is now referred
to as the Restricted Isometry Property with respect to $\ell_2$ norm (RIP-2) \cite{rip}, then the
recovery of $\ts$ from $\Phii \ts$ is indeed possible. It has been
also shown that dense random matrices constructed from Gaussian,
Bernoulli, or partial Fourier ensembles satisfy the required
RIP-2 property with high probability \cite{rip}.

Adjacency matrices of expander graphs \cite{avi} have been recently
proposed as an alternative to dense random matrices within the
compressed sensing framework, leading to computationally efficient
recovery algorithms \cite{berinde, Sina, BIR}. It has been shown that
variations of the standard recovery approaches such as \textit{basis
  pursuit} \cite{CRT} and \textit{matching pursuit} \cite{GT} are
consistent with the expander sensing approach and can recover the
original sparse signal successfully \cite{newIndyk, IR}.
In the presence of Gaussian or sparse noise, random
dense sensing and expander sensing are known to provide similar
performance in terms of the number of measurements and recovery
computation time. Berinde et al.\ proved that expander graphs with
sufficiently large expansion are near-isometries on the set of all
$k$-sparse signals in the $\ell_1$ norm; this is referred as a
Restricted Isometry Property for $\ell_1$ norm (RIP-1)
\cite{newIndyk}. Furthermore, expander sensing requires less storage
whenever the signal is sparse in the canonical basis, while random
dense sensing provides slightly tighter recovery bounds \cite{BIR}.

The approach described in this paper consists of the following key
elements:
\begin{itemize}
\item expander sensing matrices and the RIP-1 associated with them;
\item a reconstruction objective function which explicitly
incorporates the Poisson likelihood;
\item a countable collection of candidate estimators; and
\item a penalty function defined over the collection of candidates,
which satisfies the Kraft inequality and which can be used to promote
sparse reconstructions.
\end{itemize}
In general, the penalty function is selected to be small for signals
of interest, which leads to theoretical guarantees that errors are
small with high probability for such signals.  In this paper,
exploiting the RIP-1 property and the non-negativity of the
expander-based sensing matrices, we show that, in contrast to random
dense sensing, expander sensing empowered with a maximum \textit{a
  posteriori} (MAP) algorithm can approximately recover the original
signal in the presence of Poisson noise, and we prove bounds which
quantify the MAP performance.  As a result, in the presence of Poisson
noise, expander graphs not only provide general storage advantages,
but they also allow for efficient MAP recovery methods with
performance guarantees comparable to the best $k$-term approximation
of the original signal. Finally, the bounds are tighter than those for specific
dense matrices proposed by Willett and Raginsky \cite{wr, rhmw}
whenever the signal is sparse in the canonical domain, in that a log
term in the bounds in \cite{rhmw} is absent from the bounds presented
in this paper.

\subsection{Relationship with dense sensing matrices for Poisson CS}

In recent work, the authors established performance bounds for CS in
the presence of Poisson noise using dense sensing matrices based on
appropriately shifted and scaled Rademacher ensembles \cite{wr,rhmw}. Several features distinguish that work from the
present paper:
\begin{itemize}
\item The dense sensing matrices used in \cite{wr,rhmw} require more
  memory to store and more computational resources to apply to a signal
  in a reconstruction algorithm. The expander-based approach
  described in this paper, in contrast, is more efficient.
\item The expander-based approach described in this paper works {\em only} when the signal of interest is sparse
  in the canonical basis. In contrast, the dense sensing
  matrices used in \cite{wr,rhmw} can be applied to arbitrary sparsity
  bases (though the proof technique there needs to be altered slightly
  to accommodate sparsity in the canonical basis).
\item The bounds in {\em both} this paper and \cite{wr,rhmw} reflect a
  sobering tradeoff between performance and the number of measurements
  collected. In particular, more measurements (after some critical
  minimum number) can actually {\em degrade} performance as a limited
  number of events (e.g., photons) are distributed among a growing
  number of detectors, impairing the SNR of the measurements.
\end{itemize}

\subsection{Notation}

 Nonnegative reals (respectively, integers) will be denoted by $\Rplus$ (respectively, $\Zplus$). Given a vector ${u} \in \R^n$ and a set $S \subseteq
 \{1,\ldots,n\}$, we will denote by ${u}^S$ the vector obtained by
 setting to zero all coordinates of ${u}$ that are in $S^c$, the
 complement of $S$: $\forall 1 \le i \le n, u^S_i = u_i 1_{\{ i \in S
   \}}$. Given some $1 \le k \le n$, let $S$ be the set of positions
 of the $k$ largest (in magnitude) coordinates of ${u}$. Then
 ${u}^{(k)} \deq {u}^S$ will denote the {\em best $k$-term
   approximation} of ${u}$ (in the canonical basis of $\R^n$), and
$$
\kerr{{u}} \deq \| {u} - {u}^{(k)} \|_1 = \sum_{i \in S^c} |u_i|
$$
will denote the resulting $\ell_1$ approximation error. The $\ell_0$
quasinorm measures the number of nonzero coordinates of ${u}$: $\| {u}
\|_0 \deq \sum^n_{i=1} 1_{\{ u_i \neq 0 \}}$. For a subset $S
\subseteq \{1,\ldots,n\}$ we will denote by $I_S$ the vector with
components $1_{\{i \in S\}}$, $1 \le i \le n$. Given a vector ${u}$,
we will denote by ${u}^+$ the vector obtained by setting to zero all
negative components of ${u}$: for all $1 \le i \le n$, $u^+_i =
\max\{0,u_i\}$. Given two vectors $u,v \in \R^n$, we will write $u
\succeq v$ if $u_i \ge v_i$ for all $1 \le i \le n$. If $u \succeq
\alpha I_{\{1,\ldots,n\}}$ for some $\alpha \in \R$, we will simply
write $u \succeq \alpha$. We will write $\succ$ instead of $\succeq$
if the inequalities are strict for all $i$.

\subsection{Organization of the paper}

This paper is organized as follows. In Section~\ref{sec:expander}, we
summarize the existing literature on expander graphs applied to
compressed sensing and the RIP-1 property. Section~\ref{sec:poisson}
describes how the problem of compressed sensing with Poisson noise can
be formulated in a way that explicitly accounts for nonnegativity
constraints and flux preservation (i.e.,~we cannot detect more events
than have occurred); this section also contains our main theoretical result
bounding the error of a sparsity penalized likelihood reconstruction
of a signal from compressive Poisson measurements. These results are
illustrated and further analyzed in Section~\ref{sec:packets}, in
which we focus on the specific application of efficiently estimating
packet arrival rates. Several technical discussions and proofs have been relegated to the appendices.

 \section{Background on expander graphs}
 \label{sec:expander}
 
We start by defining an \textit{unbalanced bipartite vertex-expander
  graph}.
  
 \begin{definition}
\label{def:genexpand} We say that a bipartite simple graph $\Graph = (\Vin,\Vout,\Edges)$ with (regular) left degree\footnote{That is, each node in $\Vin$ has the same number of neighbors in $\Vout$.} $d$ is a $(k,\epsilon)$-{\em expander} if, for any $S \subset A$ with $|S|\leq k$,
the set of neighbors ${\cal N}(S)$ of $S$ has size $|{\cal
  N}(S)|>(1-\epsilon)d\,|S|$.
\end{definition}

\begin{figure}[!t]
\centering
\includegraphics[width=3in]{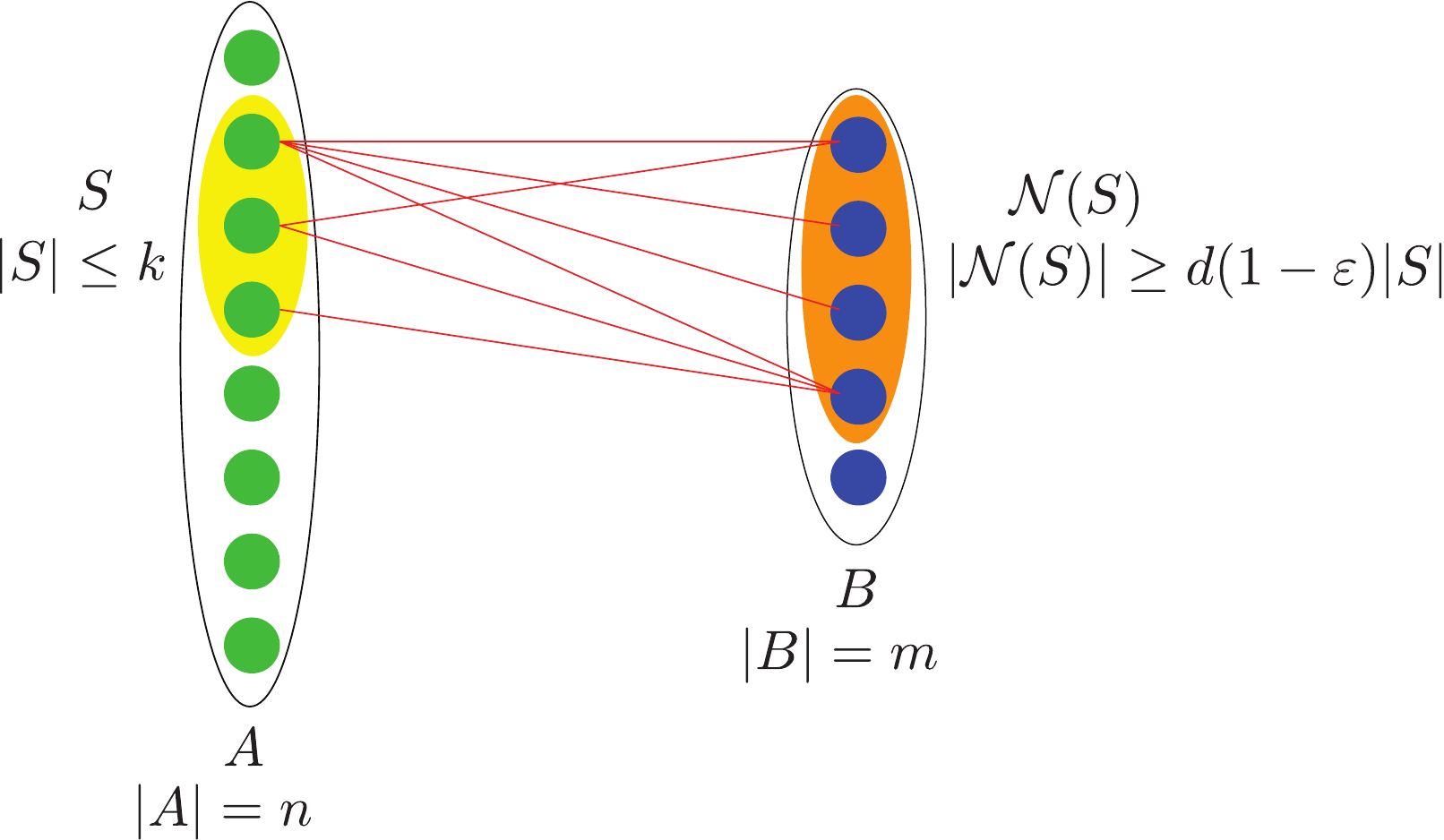}
\caption{A $(k,\epsilon)$-expander. In this example, the green
  nodes correspond to $A$, the blue nodes correspond to $B$, the yellow
  oval corresponds to the set $S \subset A$, and the orange oval corresponds to the
  set ${\cal N}(S) \subset B$. There are three colliding
  edges.}
\label{exp1}
\end{figure}

\noindent Figure~\ref{exp1} illustrates such a graph. Intuitively a
bipartite graph is an expander if any sufficiently small subset of its
variable nodes has a sufficiently large neighborhood. In the CS
setting, $\Vin$ (resp.,~$\Vout$) will correspond to the components of
the original signal (resp., its compressed representation). Hence, for
a given $|\Vin|$, a ``high-quality'' expander should have $|\Vout|$,
$d$, and $\eps$ as small as possible, while $k$ should be as close as
possible to $|\Vout|$.  The following proposition, proved using the
probabilistic method \cite{alon}, is well-known in the
literature on expanders:

\begin{proposition}[Existence of high-quality expanders]\label{prop:expander_exist}
  For any $1 \leq k \leq \frac{n}{2}$ and any  $\epsilon \in (0,1)$,
  there exists a $(k,\epsilon)$-expander with left degree
  $d=O\left(\frac{\log(n/k)}{\epsilon}\right)$ and right set
  size $m=O\left(\frac{k\log(n/k)}{\epsilon^2}\right).$
\end{proposition}

\noindent Unfortunately, there is no explicit construction of expanders from Definition~\ref{def:genexpand}. However, it can be shown that, with high probability, any $d$-regular random
graph with 
$$ d=O\left(\frac{\log(n/k)}{\epsilon}\right)\mbox{ and
  }m=O\left(\frac{k\log(n/k)}{\epsilon^2}\right)$$
satisfies the required expansion property. Moreover, the graph may be assumed to be {\em right-regular} as well, i.e., every node in $\Vout$ will have the same (right) degree $D$ \cite{SS}. Counting the number of edges in two ways, we conclude that 
$$
|\Edges|=|\Vin|d=|\Vout|D \qquad \Longrightarrow \qquad D = O\left(\frac{n}{k}\right).
$$
Thus, in practice it may suffice to use random bipartite regular graphs instead of expanders\footnote{Briefly, we can first generate a random left-regular graph with left degree $d$ (by choosing each edge independently). That graph is, with overwhelming probability, an expander graph. Then, given an expander graph which is only left-regular, a paper by Guruswami et al.~\cite{GLR} shows how to construct an expander graph with almost the same parameters, which is both left-regular and right-regular.}. Moreover, there exists an explicit construction for a class of expander graphs that comes very close to the guarantees of Proposition~\ref{prop:expander_exist}. This construction, due to Guruswami et
al.~\cite{PV2}, uses Parvaresh-Vardy codes \cite{PV} and has the following guarantees:

\begin{proposition}[Explicit construction of high-quality expanders]
\label{prop:expander_explicit}
For any positive constant $\beta$, and any $n,k,
\epsilon$, there exists a deterministic explicit construction of a
$(k,\epsilon)$-expander graph with $d=O\left({\left(\frac{\log
        n}{\epsilon}\right )}^{\frac{1+\beta}{\beta}}\right)$ and
$m=O(d^2 k^{1+\beta})$.
\end{proposition}

Expanders have been recently proposed as a means of constructing
efficient compressed sensing algorithms \cite{Sina,newIndyk,IR,
  GLR}. In particular, it has been shown that any $n$-dimensional
vector that is $k$-sparse can be fully recovered using $O\left(k\log
  \left( \frac{n}{k}\right)\right)$ measurements in $O\left(n \log
  \left( \frac{n}{k}\right)\right)$ time \cite{Sina,IR}. It has been
also shown that, even in the presence of noise in the
measurements, if the noise vector has low $\ell_1$ norm, expander-based algorithms can approximately recover any $k$-sparse
signal \cite{newIndyk, IR, BIR}. \noindent One reason why expander graphs are good sensing
candidates is that the adjacency matrix of any $(k,\eps)$-expander almost
preserves the $\ell_1$ norm of any $k$-sparse vector \cite{newIndyk}. In
other words, if the adjacency matrix of an expander is used for
measurement, then the $\ell_1$ distance between two
sufficiently sparse signals is preserved by measurement.  This
property is known as the ``Restricted Isometry Property for $\ell_1$
norms'' or the ``RIP-1'' property. Berinde et al.
have shown that this condition is sufficient for sparse recovery using
$\ell_1$ minimization \cite{newIndyk}.

The precise statement of the RIP-1 property, whose proof can be found in \cite{Sina}, goes as follows:

\begin{lemma}[RIP-1 property of the expander graphs]
\label{RIP}
Let $\AdjMat$ be the $m \times n$ adjacency matrix of a $(k,\epsilon)$
expander graph $G$. Then for any $k$-sparse vector $x\in \mathbb{R}^n$
we have:
\begin{equation}
\label{ripeq}
(1-2\epsilon)d \|x\|_1  \leq \|\AdjMat x\|_1 \leq d \|x\|_1
\end{equation}
\end{lemma}

The following proposition is a direct consequence of the above RIP-1
property. It states that if, for any almost $k$-sparse
vector\footnote{By ``almost sparsity'' we mean that the vector has at
  most $k$ significant entries.}  $u$, there exists a vector $v$
whose $\ell_1$ norm is close to that of $u$, and if $\AdjMat v$ approximates
$\AdjMat u$, then $v$ also approximates $u$. Our results of
Section~\ref{sec:poisson} exploit the fact that the proposed MAP decoding
algorithm outputs a vector satisfying the two conditions above, and
hence approximately recovers the desired signal.
\begin{proposition}
\label{piotr}
Let $\AdjMat$ be the adjacency matrix of a $(2k,\epsilon)$-expander and
$u,v$ be two vectors in $\mathbb{R}^n$, such
that
$$\|u\|_1\geq  \|v\|_1-\Delta$$
for some $\Delta > 0$. Then $\| u - v \|_1$ is upper-bounded by
$$ 
\|u-v\|_1 \leq \frac{1-2\epsilon}{1-6\epsilon}\left(2\kerr{u}+\Delta\right)
 + \frac{2}{d(1-6\epsilon)}\|\AdjMat u- \AdjMat v\|_1. 
$$
In particular, if we let $\epsilon = 1/16$, then we get the bound
$$
\| u - v \|_1 \leq 4\kerr{u} + \frac{4}{d} \| \AdjMat u - \AdjMat v \|_1 + 2 \Delta.
$$
\end{proposition}
\begin{IEEEproof}See Appendix~\ref{app:piotr}.\end{IEEEproof}

For future convenience, we will introduce the following piece of notation. Given $\inDim$ and $1 \le k \le \inDim/4$, we will denote by $\Graph_{k,\inDim}$ a $(2k,1/16)$-expander with left set size $\inDim$ whose existence is guaranteed by Proposition~\ref{prop:expander_exist}. Then $\Graph_{k,\inDim} = (\Vin,\Vout,\Edges)$ has
$$
|\Vin| = \inDim, \quad |\Vout| = \outDim = O(k \log (n/k)), \quad d = O(\log (n/k)).
$$

 \section{Compressed sensing in the presence of Poisson Noise}
 \label{sec:poisson}
 
 \subsection{Problem statement}
 
 We wish to recover an unknown vector $\ts \in \Rplus^\inDim$ of Poisson intensities from a measured vector $y \in \Zplus^\outDim$, sensed according to the Poisson model
\begin{align}
y \sim \text{Poisson}(\Phi \ts), \label{eq:obs}
\end{align}
where $\Phi \in \Rplus^{\outDim \times \inDim} $ is a
positivity-preserving sensing matrix\footnote{Our choice of this
    observation model as opposed to a ``shot-noise'' model based on
    $\Phi$ operating on Poisson observations of $\ts$ is discussed in
    Appendix~\ref{app:shot}.}. That is, for each $j \in
\{1,\ldots,\outDim\}$, $y_j$ is sampled independently from a Poisson
distribution with mean $(\Phi\ts)_j$:
\begin{equation}
\Prob_{\Phi\ts}(y) = \prod^m_{j=1} \Prob_{(\Phi\ts)_j}(y_j),
\end{equation}
where, for any $z \in \Zplus$ and $\lambda \in \Rplus$, we have
\begin{equation}
\label{poisson}
\Prob_{\lambda}(z) \deq
\begin{cases}
	\displaystyle\frac{\lambda^z}{z!}e^{-\lambda}& \mbox{if } \lambda > 0
	\\ 1_{\{z = 0\}} & \mbox{otherwise}
\end{cases},
\end{equation}
where the $\lambda = 0$ case is a consequence of the fact that
$$\lim_{\lambda\rightarrow 0}
\frac{\lambda^z}{z!}e^{-\lambda}=1_{\{z = 0\}}.$$
We assume that the $\ell_1$ norm of $\ts$ is known, $\| \ts \|_1 = L$ (although later we will show that this assumption can be relaxed). We are interested in designing a sensing matrix $\Phi$ and an estimator $\that = \that(y)$, such that $\ts$ can be recovered with small expected $\ell_1$ risk
$$
\Risk{\that}{\ts} \deq \Exp_{\Phi \ts} \| \that - \ts \|_1,
$$
where the expectation is taken w.r.t.\ the distribution $\Prob_{\Phi \ts}$.

\subsection{The proposed estimator and its performance}
\label{subsec:theory}
To recover $\ts$, we will use a penalized Maximum Likelihood
Estimation (pMLE) approach. Let us choose a convenient $1 \le k \le
\inDim/4$ and take $\Phi$ to be the normalized adjacency matrix of the
expander $\Graph_{k,\inDim}$ (cf.~Section~\ref{sec:expander} for
definitions): $\nAdjMat \deq \AdjMat/d$. Moreover, let us choose a
finite or countable set $\Theta_L$ of candidate estimators $\tb \in
\Rplus^\inDim$ with $\|\tb\|_1 \le L$, and a {\em penalty} $\pen :
\Theta_L \to \Rplus$ satisfying the {\em Kraft
  inequality}\footnote{Many penalization functions can be modified
  slightly (e.g., scaled appropriately) to satisfy the Kraft
  inequality. All that is required is a finite collection of
  estimators (i.e., $\Theta_L$) and an associated prefix code for each
  candidate estimate in $\Theta_L$. For instance, this would certainly
  be possible for a total variation penalty, though the details are
  beyond the scope of this paper.}
\begin{equation}
\label{eq:kraft}
\sum_{\tb \in \Theta_L} e^{-\pen(\tb)} \le 1.
\end{equation}
For instance, we can impose less penalty on sparser signals or
construct a penalty based on any other prior knowledge about the
underlying signal. 

With these definitions, we consider the following {\em penalized
    maximum likelihood estimator (pMLE)}:
\begin{align}\label{eq:PMLE_main}
  \that &\deq \argmin_{\tb \in \Theta_L} \left[ - \log \Prob_{\nAdjMat \tb}
    (\yb) + 2\, \pen(\tb) \right] 
\end{align}
\noindent One way to think about the procedure in (\ref{eq:PMLE_main}) is as a Maximum \post Probability (MAP) algorithm over the
set of estimates $\Theta_L$, where the likelihood is computed according to
the Poisson model \eqref{poisson} and the penalty function corresponds
to a negative log prior on the candidate estimators in $\Theta_L$.

Our main bound on the performance of the pMLE is as follows:

\begin{theorem}
\label{thm:expander2_main}
Let $\A$ be the normalized adjacency matrix of $\Graph_{k,\inDim}$, let  $\ts\in\Rplus^\inDim$ be the original signal
compressively sampled in the presence of Poisson noise, and let $\that$ be obtained through \eqref{eq:PMLE_main}. Then
\begin{align}
&\Risk{\that}{\ts} \le 4 \kerr{\ts} \nonumber \\
& \qquad + 8 \sqrt{L \min_{\tb \in \Theta_L} \left[
    \RE(\Prob_{\nAdjMat\ts}\parallel
      \Prob_{\nAdjMat\tb})+2\, \pen(\tb)\right]},
\label{eq:pMLE_bound_main}
\end{align}
where
$$
\RE(\Prob_{g}\|\Prob_{h}) \deq \sum_{y \in \Zplus^m} \Prob_{g}(y)\log \frac{\Prob_{g}(y)}{\Prob_{h}(y)}
$$
is the Kullback--Leibler divergence (relative entropy) between $\Prob_g$ and $\Prob_h$ \cite{CoverThomas}.
\end{theorem}
\begin{IEEEproof}
Since $\that \in \Theta_L$, we have $L = \| \ts \|_1 \ge \| \that \|_1$. Hence, using Proposition~\ref{piotr} with $\Delta = 0$, we can write
\begin{align*}
\| \ts - \that \|_1 \le 4 \kerr{\ts} + 4 \| \nAdjMat (\ts - \that) \|_1.
\end{align*}
Taking expectations, we obtain
\begin{align}
\Risk{\that}{\ts} &\le 4 \kerr{\ts} + 4 \Exp_{\nAdjMat \ts} \| \nAdjMat (\ts - \that) \|_1 \nonumber \\ 
&\le 4 \kerr{\ts} + 4 \sqrt{\Exp_{\nAdjMat \ts} \| \nAdjMat(\ts - \that) \|^2_1} \label{eq:pMLE_risk_main}
\end{align}
where the second step uses Jensen's inequality. Using Lemmas~\ref{l1} and \ref{l2} in Appendix~\ref{app:lemmas}, we have
\begin{align*}
 \Exp_{\nAdjMat \ts} \| \nAdjMat(\ts - \that) \|^2_1 \le 4L \min_{\tb \in \Theta_L} \left[
    \RE(\Prob_{\nAdjMat\ts}\parallel
      \Prob_{\nAdjMat\tb})+2\, \pen(\tb)\right]
\end{align*}
Substituting this into \eqref{eq:pMLE_risk_main}, we obtain
\eqref{eq:pMLE_bound_main}.
\end{IEEEproof}

The bound of Theorem~\ref{thm:expander2_main} is an {\em oracle inequality}: it states that the $\ell_1$ error of $\that$ is (up to multiplicative constants) the sum of the $k$-term approximation error of $\ts$ plus $\sqrt{L}$ times the minimum penalized relative entropy error over the set of candidate estimators $\Theta_L$. The first term in \eqref{eq:pMLE_bound_main} is smaller for sparser
$\ts$, and the second term is smaller when there is a $\tb \in \Theta_L$ which is
simultaneously a good approximation to $\ts$ (in the sense that the distributions $\Prob_{\Phi \ts}$ and $\Prob_{\Phi \tb}$ are close) and has a low penalty.

\begin{remark}\label{rem:L1_estimation}
So far we have assumed that the $\ell_1$ norm of $\ts$ is known \textit{a priori}. If this is not the case, we can still estimate it with high accuracy using noisy compressive measurements. Observe that, since each measurement $y_j$ is a Poisson random variable with mean $\left(\Phi \theta^*\right)_j$, $\sum_j y_j$ is Poisson with mean $\| \Phi \theta^* \|_1$. Therefore, $\sqrt{\sum_j y_j}$ is approximately normally distributed with mean 
$\approx \sqrt{\| \Phi \theta^* \|_1}$ and variance $\approx \frac{1}{4}$ \cite[Sec.~6.2]{stabilized}.\footnote{This observation underlies the use of variance-stabilizing transforms.} Hence, Mill's inequality \cite[Thm.~4.7]{Wasserman} guarantees that, for every positive $t$,
$$
\Pr\left[ \left|\sqrt{\sum_j y_j}- \sqrt{\| \Phi \theta^* \|_1}\right| > t\right] \lesssim \frac{e^{-2t^2}}{\sqrt{2\pi} t},
$$ 
where $\lesssim$ is meant to indicate the fact that this is only an approximate bound, with the approximation error controlled by the rate of convergence in the central limit theorem. Now we can use the RIP-1 property of the expander graphs obtain the estimates
$$\left(\sqrt{\sum_j y_j}- t\right)^2\leq  \| \Phi \theta^* \|_1 \leq \|\theta^*\|_1,$$ 
and
$$\frac{\left(\sqrt{\sum_j y_j}+ t\right)^2}{(1-2\epsilon)}\geq  \frac{\| \Phi \theta^* \|_1}{(1-2\epsilon)} \geq \|\theta^*\|_1$$
that hold with (approximate) probability at least $1-(\sqrt{2\pi}t)^{-1}e^{-2t^2}$.
\end{remark}

\subsection{A bound in terms of $\ell_1$ error}

The bound of Theorem~\ref{thm:expander2_main} is not always useful since it bounds the $\ell_1$ risk of the pMLE in terms of the relative entropy. A bound purely in terms of $\ell_1$ errors would be more desirable. However, it is not easy to obtain without imposing extra conditions either on $\ts$ or on the candidate estimators in $\Theta_L$. This follows from the fact that the divergence $\RE(\Prob_{\Phi \theta^*} \| \Prob_{\Phi \theta})$ may take the value $+\infty$ if there exists some $y$ such that $\Prob_{\Phi \theta}(y) = 0$ but $\Prob_{\Phi \theta^*}(y) > 0$.

One way to eliminate this problem is to impose an additional requirement on the candidate estimators in $\Theta_L$: There exists some $c > 0$, such that
\begin{align}\label{eq:strictly_positive}
	\Phi \tb \succeq c, \qquad \forall \tb \in \Theta_L
\end{align}
Under this condition, we will now develop a risk bound for the pMLE purely in terms of the $\ell_1$ error.

\begin{theorem}
\label{thm:expander2}
Suppose that all the conditions of Theorem~\ref{thm:expander2_main} are satisfied. In addition, suppose that the set $\Theta_L$ satisfies the condition \eqref{eq:strictly_positive}. Then
\begin{align}
&\Risk{\that}{\ts} \le 4 \kerr{\ts} + 8 \sqrt{L \min_{\tb \in \Theta_L} \left[ \frac{\| \ts - \tb \|^2_1 }{c} +  2\,\pen(\tb)\right]}.
\label{eq:pMLE_bound_6}
\end{align}
\end{theorem}
\begin{IEEEproof}
Using Lemma~\ref{l3} in Appendix~\ref{app:lemmas}, we get the bound
$$
	\RE(\Prob_{\Phi \ts} \| \Prob_{\Phi \theta}) \le \frac{1}{c} \| \ts - \tb \|^2_1, \qquad \forall \tb \in \Theta_L.
$$
Substituting this into Eq.~\eqref{eq:pMLE_bound_main}, we get \eqref{eq:pMLE_bound_6}.
\end{IEEEproof}

\begin{remark}Because every $\tb \in \Theta_L$ satisfies $\| \tb \|_1 \le L$, the constant $c$ cannot be too large. In particular, if \eqref{eq:strictly_positive} holds, then for every $\tb \in \Theta_L$ we must have
\begin{align*}
	\| \Phi \tb \|_1 \ge \outDim \min_j (\Phi \tb)_j \ge \outDim c.
\end{align*}
On the other hand, by the RIP-$1$ property we have $\| \Phi \tb \|_1 \le \| \tb \|_1 \le L$. Thus, a necessary condition for \eqref{eq:strictly_positive} to hold is $c \le L/m$. Since $m = O(k \log (n/k))$, the best risk we may hope to achieve under some condition like \eqref{eq:strictly_positive} is on the order of
\begin{align}
	&\Risk{\that}{\ts} \le 4 \kerr{\ts} \nonumber \\
	& \qquad \qquad  + C \sqrt{ \min_{\tb \in \Theta_L} \left[ k \log(n/k) \| \tb - \ts \|^2_1 +  L\, \pen(\tb)\right]}
	\label{eq:pMLE_bound_6a}
\end{align}
for some constant $C$, e.g., by choosing $c \propto \frac{L}{k\log (n/k)}$. Effectively, this means that, under the positivity condition \eqref{eq:strictly_positive}, the $\ell_1$ error of $\that$ is the sum of the $k$-term approximation error of $\ts$ plus $\sqrt{m} = \sqrt{k \log (n/k)}$ times the best penalized
$\ell_1$ approximation error. The first term in \eqref{eq:pMLE_bound_6a} is smaller for sparser
$\ts$, and the second term is smaller when there is a $\tb \in \Theta_L$ which is
simultaneously a good $\ell_1$ approximation to $\ts$ and has a low penalty.
\end{remark}

 \subsection{Empirical performance}
 
 Here we present a simulation study that validates our 
 method.  In this experiment, compressive Poisson observations are
 collected of a randomly generated sparse signal passed through the 
 sensing matrix generated from an adjacency matrix of an expander.
 We then reconstruct the signal by utilizing an algorithm that
 minimizes the objective function in \eqref{eq:PMLE_main}, and assess the
 accuracy of this estimate.  We repeat this procedure over several
 trials to estimate the average performance of the method.

 More specifically, we generate our length-$n$ sparse signal $\ts$
 through a two-step procedure.  First we select $k$ elements of $\{1,\ldots,n\}$ uniformly
 at random, then we assign these elements an intensity $I$.  All other
 components of the signal are set to zero.  For these experiments, we
 chose a length $n=\text{100,000}$ and varied the sparsity $k$ among three
 different choices of 100, 500, and 1,000 for two intensity levels $I$ of 10,000 
 and 100,000. We then vary the number $m$ of Poisson observations from 100 to 20,000
 using an expander graph sensing matrix with degree $d = 8$. Recall that
 the sensing matrix is normalized such that the total signal intensity is divided
 amongst the measurements, hence the seemingly high choices of $I$.

To reconstruct the signal, we utilize the SPIRAL-$\ell_1$ algorithm
\cite{harmany:PCS} which solves \eqref{eq:PMLE_main} when $\pen(\tb) =
\tau \|\tb\|_1$.  We design the algorithm to optimize 
over the continuous domain $\reals_+^n$
instead of the discrete set $\Theta_L$.  This is equivalent to the proposed
pMLE formulation in the limit as the discrete set of 
estimates becomes increasingly dense in the set of all $\theta \in \reals^n_+$ with $\| \theta \|_1 \le L$, i.e., we quantize this set on an ever finer scale, increasing the bit allotment to represent each $\theta$. In this high-resolution limit, the Kraft inequality requirement \eqref{eq:kraft} on the penalty $\pen(\theta)$ will translate to $\int e^{-\pen(\theta)} d\theta < \infty$.  If we select a penalty proportional to the negative log of a prior
probability distribution for $\theta$, this requirement will be satisfied. From a Bayesian perspective,
the $\ell_1$ penalty arises by assuming each component $\theta_i$ is drawn i.i.d.\ from a zero-mean Laplace prior $p(\theta_i) = e^{-|\theta_i|/b}/2b$.  Hence the regularization parameter $\tau$ is inversely related to the scale parameter $b$ of the prior, as a larger $\tau$ (smaller $b$) will promote solutions with more zero-valued components.

This relaxation results in a computationally tractable convex program over a continuous domain, albeit implemented on a machine with finite precision. The SPIRAL algorithm utilizes a sequence of quadratic
subproblems derived by using a second-order Taylor expansion of the
Poisson log-likelihood at each iteration.  These subproblems are made
easier to solve by using a separable approximation whereby the
second-order Hessian matrix is approximated by a scaled identity
matrix.  For the particular case of the $\ell_1$ penalty, these
subproblems can be solved quickly, exactly, and noniteratively by a
soft-thresholding rule.  

\begin{figure}
\begin{center}
\includegraphics[width=\columnwidth]{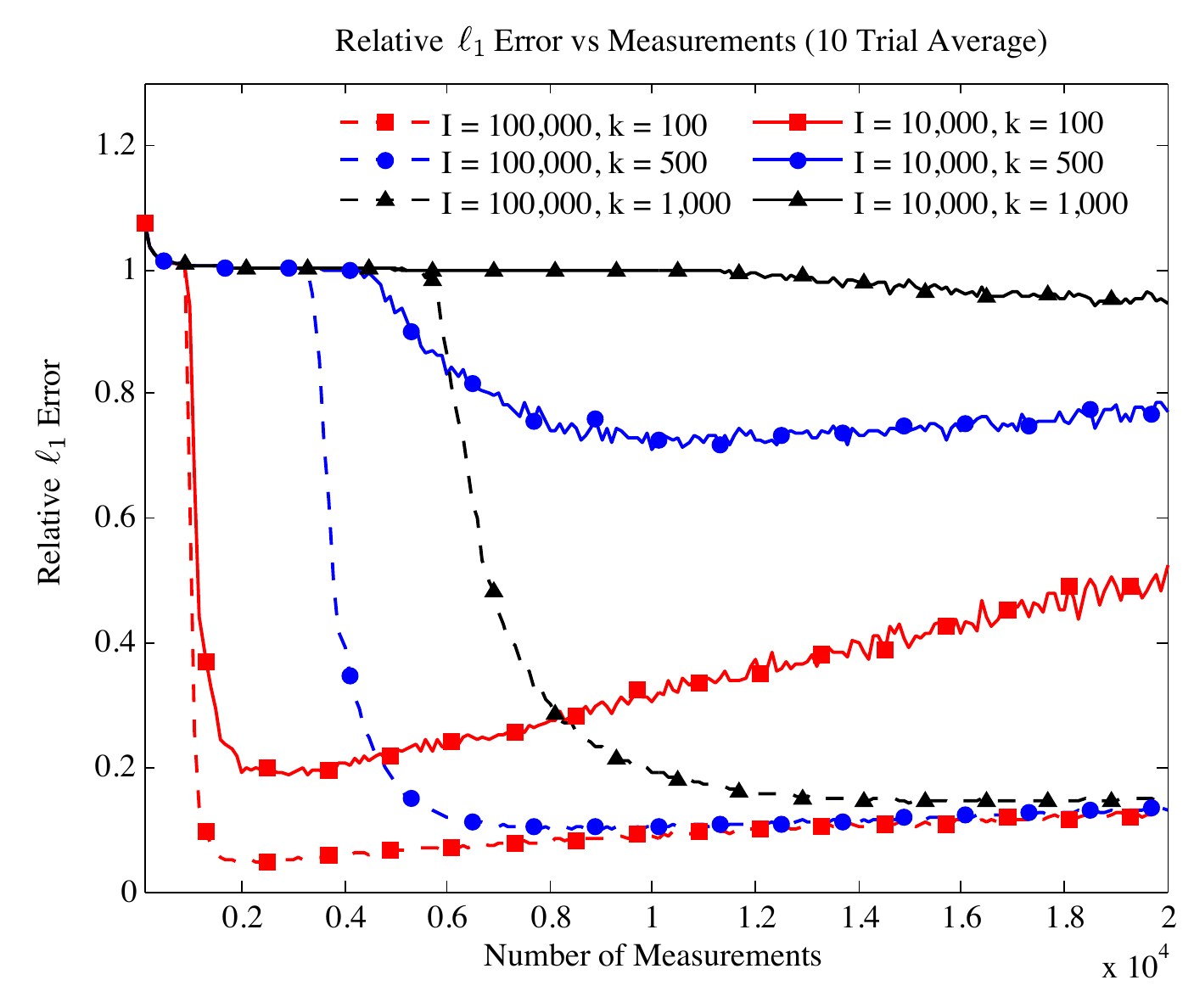}
\end{center}
\caption{Average performance (as measured by the normalized $\ell_1$
  error $\|\ts - \that\|_1/ \|\ts\|_1$) for the proposed expander-based
  observation method for recovering sparse signals under Poisson
  noise.  In this experiment, we sweep over a range of measurements
  and} consider a few sparsity ($k$) and intensity $(I)$ levels of the true signal.
\label{fig:results}
\end{figure}

After reconstruction, we assess the estimate $\that$ according to the
normalized $\ell_1$ error $\|\ts - \that\|_1/ \|\ts\|_1$.  We select
the regularization weighting $\tau$ in the SPIRAL-$\ell_1$ algorithm to minimize
this quantity for each randomly 
generated experiment indexed by $(I,k,m)$. To assure
that the results are not biased in our favor by only considering a
single random experiment for each $(I,k,m)$, we repeat this experiment several times. The
averaged reconstruction accuracy over 10 trials is presented in
Figure~\ref{fig:results}. 

These results show that the proposed method is able to accurately
estimate sparse signals when the signal intensity is sufficiently
high; however, the performance of the method degrades for lower
signal strengths. More interesting is the behavior as we vary the
number of measurements.  There is a clear phase transition where accurate
signal reconstruction becomes possible, however the performance gently degrades
with the number of measurements since there is a lower signal-to-noise ratio 
per measurement.  This effect is more pronounced at lower intensity levels, 
as we more quickly enter the regime where only a few photons are collected 
per measurement. These findings support the error bounds developed in
Section~\ref{subsec:theory}. 

\section{Application: Estimating packet arrival rates}
\label{sec:packets}

This section describes an application of the pMLE estimator of Section~\ref{sec:poisson}:   an indirect approach  for
reconstructing average  packet arrival rates  and instantaneous packet
counts for a given number of streams (or flows) at a  router in a communication network,  where the arrivals of
packets  in each  flow are assumed to follow  a Poisson  process. All packet counting must be done in hardware at the router, and
any hardware implementation must strike a delicate balance between speed, accuracy, and cost. For instance, one  could  keep a
dedicated counter for each flow,  but, depending on the type of memory
used, one could end up with  an implementation that is either fast but
expensive  and  unable  to keep  track  of  a  large number  of  flows
(e.g.,~using SRAMs, which have low access times, but are expensive and
physically  large) or  cheap  and high-density  but slow  (e.g.,~using
DRAMs, which are cheap and small, but have longer access times) \cite{ElephantsMice,CounterBraids}.

However,          there         is          empirical         evidence
\cite{EmpiricalFlow1,EmpiricalFlow2}  that flow  sizes in  IP networks
follow  a {\em  power-law} pattern:  just a  few flows  (say, $10\%$)
carry most  of the traffic  (say, $90\%$). Based on  this observation,
several investigators have proposed methodologies for estimating flows
using a small  number of counters by either (a)  keeping track only of
the flows whose  sizes exceed a given fraction  of the total bandwidth
(the approach suggestively termed ``focusing on the elephants, ignoring the mice") \cite{ElephantsMice} or  (b) using  sparse random graphs  to aggregate
the  raw packet  counts  and  recovering flow  sizes  using a  message
passing decoder \cite{CounterBraids}.

We consider an alternative to these approaches based on Poisson CS, assuming that the underlying Poisson rate vector is sparse or approximately
sparse --- and, in fact, it is the approximate sparsity of the rate vector that mathematically describes the power-law behavior of the average packet counts. The goal is to maintain a compressed summary of the process
sample paths using a small number of counters, such that it is possible to reconstruct both the total number of packets in each
flow and the underlying rate vector. Since we are dealing here with Poisson streams, we would like to push the metaphor further and say that we are ``focusing on the whales, ignoring the minnows."

\subsection{Problem formulation}

We wish to monitor a large number $\inDim$ of packet flows using a
much smaller number $\outDim$ of counters. Each flow is a homogeneous
Poisson process (cf.~\cite{BertsekasNetworks} for details pertaining to Poisson processes and networking applications). Specifically, let $\Rate^* \in \Rplus^\inDim$ denote
the vector of rates, and let $U$ denote the random process
$U = \{U_t\}_{t \in \Rplus}$ with sample paths in
$\Zplus^\inDim$, where, for each $i \in \{1,\ldots,\inDim\}$, the $i$th
 component of $U$ is a
 homogeneous Poisson process with the rate of $\Rate_i$ arrivals
 per unit time, and all the component processes are mutually conditionally
 independent given $\Rate$.

 The goal is to estimate the unknown rate vector $\Rate$ based on
 $\Cts$. We will focus on performance bounds for power-law network traffic, i.e.,~for $\Rate^*$ belonging to the class
\begin{align}\label{eq:heavy_tail}
  \Sigma_{\alpha,L_0} \deq \left\{ \Rate \in \Rplus^\inDim : \| \Rate
    \|_1 = L_0; {\kerr{\Rate} = O( k^{-\alpha})} \right\}
\end{align}
for some $L_0 > 0$ and $\alpha \ge 1$, where the constant hidden in the $O(\cdot)$ notation may depend on $L_0$. Here, $\alpha$ is the power-law
exponent that controls the tail behavior; in particular, the extreme
regime $\alpha \to + \infty$ describes the fully sparse setting. As in Section~\ref{sec:poisson}, we assume the total arrival rate $\| \Rate^* \|_1$ to be known (and equal to a given $L_0$) in advance, but this assumption can be easily dispensed with (cf.~Remark~\ref{rem:L1_estimation}).

As before, we evaluate each candidate estimator $\hRate = \hRate(y)$ based on its expected $\ell_1$ risk,
$$
\Risk{\hRate}{\Rate^*} = \Exp_{\Rate^*} \| \hRate - \Rate^* \|_1.
$$

\subsection{Two estimation strategies}

We consider two estimation strategies. In both cases, we let our
measurement matrix $\AdjMat$ be the adjacency matrix of the expander
$\Graph_{k,\inDim}$ for a fixed $k \le \inDim/4$ (see Section~\ref{sec:expander} for definitions). The first strategy,
which we call the {\em direct method}, uses standard expander-based CS to construct an
estimate of $\Rates$. The second is the pMLE strategy, which relies on the
machinery presented in Section~\ref{sec:poisson} and can
be used when only the rates are of interest. 

\subsubsection{The direct method}

In this method, which will be used as a ``baseline'' for assessing the performance of the pMLE, the counters
  are updated in discrete time, every $\res$ time units. Let $\Flow =
  \{\Flow_\samp\}_{\samp \in \Zplus}$ denote the sampled version of
  $U$, where $\Flow_\samp \deq U_{\samp\res}$. The update takes place
  as follows. We have a binary matrix $\AdjMat \in \{0,1\}^{\outDim
    \times \inDim}$, and at each time $\samp$ let $\Cts_\samp =
  \AdjMat\Flow_\samp$. In other words, $\Cts$ is obtained by passing a sampled $\inDim$-dimensional homogeneous Poisson process with rate vector $\Rate$ through a linear transformation $\AdjMat$.
 
 The direct method uses expander-based CS to obtain an estimate
$\hFlow_\samp$ of $\Flow_\samp$ from $\Cts_\samp = \AdjMat \Flow_\samp$, followed by
letting
\begin{align}\label{eq:direct}
\hrate_\samp^\dir = \frac{\hFlow^+_\samp}{\samp\res}.
\end{align}
This strategy is based on the observation that $\Flow_\samp/(\samp\res)$ is
the maximum-likelihood estimator of $\Rates$. To
obtain $\hFlow_\samp$, we need to solve the convex program
\begin{align*}
\text{minimize } & \| u \|_1 \qquad \text{ subject to }  \AdjMat u = \Cts_\samp
\end{align*}
which can be cast as a linear program \cite{BerindeExpanders}. The
resulting solution $\hFlow_\samp$ may have negative
coordinates,\footnote{Khajehnejad et al.~\cite{Khaj} have recently
  proposed the use of perturbed adjacency matrices of expanders to
  recover nonnegative sparse signals.} hence the use of the
$(\cdot)^+$ operation in \eqref{eq:direct}. We then have the following
result:

\begin{theorem}\label{thm:direct}
\begin{align}\label{eq:direct2}
\Risk{\hRate_\samp^\dir}{\Rates} \le 4 \kerr{\Rates} + \frac{ \| (\Rates)^{1/2} \|_1}{\sqrt{\samp\res}},
\end{align}
where $(\Rates)^{1/2}$ is the vector with components
$\sqrt{\Rate^*_i}, \forall i$.
\end{theorem}

\begin{remark} Note that the error term in \eqref{eq:direct2} is $O(1/\sqrt{\samp})$, assuming everything else is kept constant, which coincides with the optimal rate of the $\ell_1$ error decay in parametric estimation problems.
\end{remark}

\begin{IEEEproof} We first observe that, by construction, $\hFlow_\samp$
  satisfies the relations $\AdjMat \hFlow_\samp = \AdjMat \Flow_\samp$ and $\|
  \hFlow_\samp \|_1 \le \| \Flow_\samp \|_1$. Hence,
\begin{align}
\Exp \| \hFlow_\samp - \samp\res \Rates \|_1 &\le \Exp \| \hFlow_\samp - \Flow_\samp \|_1 + \Exp \| \Flow_\samp - \samp\res \Rates \|_1 \nonumber \\
& \le 4 \Exp \kerr{\Flow_\samp} +  \Exp \| \Flow_\samp - \samp\res \Rates \|_1 \label{eq:1s02}
\end{align}
where the first step uses the triangle inequality, while the second
step uses Proposition~\ref{piotr} with $\Delta = 0$. To bound
the first term in \eqref{eq:1s02}, let $S \subset \{1,\ldots,\inDim\}$
denote the positions of the $k$ largest entries of
$\Rates$. Then, by definition of the best $k$-term
representation,
\begin{align*}
\kerr{\Flow_\samp} \le \| \Flow_\samp - \Flow^S_\samp \|_1 = \sum_{i \in S^c} |\flow_{\samp,i}| = \sum_{i \in S^c} \flow_{\samp,i}.
\end{align*}
Therefore,
\begin{align*}
\Exp \kerr{\Flow_\samp} \le \Exp \left[ \sum_{i \in S^c} \flow_{\samp,i} \right] = \samp\res \sum_{i \in S^c} \rate_i^* \equiv \samp\res \kerr{\Rates}.
\end{align*}
To bound the second term, we can use concavity of the square root, as
well as the fact that each $\flow_{\samp,i} \sim
\Poi(\samp\res\rate^*_i)$, to write
\begin{align*}
 \Exp \| \Flow_\samp - \samp\res \Rates \|_1 
&= \Exp \left[ \sum^N_{i=1} |\flow_{n,i} - \samp\res \rate^*_i | \right] \\
 &= \Exp \left[ \sum^N_{i=1} \sqrt{(\flow_{n,i} - \samp\res \rate^*_i )^2} \right] \\
&  \le \sum^\inDim_{i=1} \sqrt{ \Exp(\flow_{\samp,i} - \samp\res \rate^*_i)^2 } =  \sum^\inDim_{i=1} \sqrt{\samp\res\rate^*_i}.
\end{align*}
Now, it is not hard to show that $\| \hFlow^+_\samp - \samp\res\rate^* \|_1 \le \| \hFlow_\samp - \samp\res\rate^* \|_1$. Therefore,
\begin{align*}
\Risk{\hRate_\samp^\dir}{\Rates} &\le \frac{ \Exp \| \hFlow_\samp - \samp\res \Rates \|_1}{\samp\res} \le 4 \kerr{\Rates} + \frac{ \| (\Rates)^{1/2} \|_1}{\sqrt{\samp\res}},
\end{align*}
which proves the theorem.
\end{IEEEproof}

\subsubsection{The penalized MLE approach}

In the penalized MLE approach the counters are updated in a
  slightly different manner. Here the counters are still updated in
  discrete time, every $\res$ time units; however, each counter $i \in
  \left\{1,\cdots,m\right\}$ is updated at times $\left(\samp
    \res+\frac{i}{m}\res \right)_{\samp\in\mathbb{Z}_+}$, and only
  aggregates the packets that have arrived during the time period
  $\left[ \samp\res+\frac{i-1}{m}\res,
    \samp\res+\frac{i}{m}\res\right)$. Therefore, in contrast to the
  direct method, here each arriving packet is registered by at most
  one counter. Furthermore, since the packets arrive according to a
  homogeneous Poisson process, conditioned on the vector $\Rates$, the
  values measured by distinct counters are independent\footnote{The
    independence follows from the fact that if $X_1,\cdots,X_m$ are
    conditionally independent random variables, then for any choice of
    functions $g_1,\cdots,g_m$, the random variables
    $g_1(X_1),\cdots,g_m(X_m)$ are also conditionally independent.}. Therefore, the vector of counts at time $\samp$ obeys
$$
y_\samp \sim \text{Poisson}(\nAdjMat \ts) \qquad \text{where} \qquad  \ts = \frac{\samp\res d}{m} \Rate^*
$$
which is precisely the sensing model we have analyzed in Section~\ref{sec:poisson}.

Now assume
that the total average arrival
rate $\| \Rates\|_1 = L_0$ is known. Let $\Lambda$ be a finite or a countable set of candidate estimators with $\| \Rate \|_1 \le L_0$ for
all $\Rate \in \Lambda$, and let $\pen(\cdot)$ be a penalty functional
satisfying the Kraft inequality over $\Lambda$. Given $\samp$ and $\res$,
consider the scaled set
$$
\Lambda_{\samp,\res} \deq \frac{\samp\res d}{m} \Lambda \equiv \left\{ \frac{\samp\res d}{m} \Rate : \Rate \in \Lambda\right\}
$$
with the same penalty
function, $\pen\left(\frac{\samp\res d}{m} \Rate\right) = \pen(\Rate)$ for all $\Rate \in \Lambda$. We can now apply the results of Section~\ref{sec:poisson}. Specifically, let
$$
\hRate^\pMLE_\samp \deq \frac{m\,\that}{\samp\res d},
$$
where $\that$ is the corresponding pMLE estimator obtained according to \eqref{eq:PMLE_main}. The following theorem is a consequence of Theorem~\ref{thm:expander2} and the remark following it: 

\begin{theorem}\label{thm:pMLE} If the set $\Lambda$ satisfies the strict positivity condition \eqref{eq:strictly_positive}, then there exists some absolute constant $C > 0$, such that
\begin{align}
&\Risk{\hRate^\pMLE_\samp}{\Rates} \le 4\kerr{\Rates} \nonumber\\
& \qquad + C \sqrt{\min_{\Rate \in \Lambda} \left[ k \log(\inDim/k) \| \Rate - \Rate^* \|^2_1 + \frac{k\, L_0\, \pen(\Rate)}{\samp \res} \right]}.
\label{eq:pMLE_bound}
\end{align}
\end{theorem}
We now develop risk bounds under the power-law condition. To this
end, let us suppose that $\Rates$ is a member of the power-law
class $\Sigma_{L_0,\alpha}$ defined in \eqref{eq:heavy_tail}. Fix a
small positive number $\delta$, such that $L_0/\sqrt{\delta}$ is an
integer, and define the set
\begin{align*}
 \Lambda \deq \left\{ \Rate \in \Rplus^\inDim: \| \Rate \|_1 \le L_0;  \rate_i \in \{s \sqrt{\delta}\}^{L_0/\sqrt{\delta}}_{s=0}, \forall i \right\}
\end{align*}
These will be our candidate estimators of $\Rates$. We can define
the penalty function $\pen(\Rate) \asymp \| \Rate \|_0 \log
(\delta^{-1})$. For any $\Rate \in \Sigma_{\alpha,L_0}$ and
any $1 \le r \le \inDim$ we can find some $\Rate^{(r)} \in
\Lambda$, such that $\| \Rate^{(r)} \|_0 \asymp r$ and
$$
\| \Rate - \Rate^{(r)} \|^2_1 \asymp r^{-2\alpha} + r\,\delta.
$$
Here we assume that $\delta$ is sufficiently small, so that the
penalty term $\frac{k\,r \log (\delta^{-1})}{\samp\res}$ dominates the
quantization error $r\,\delta$. In order to guarantee that the penalty function satisfies Kraft's inequality, we need to ensure that
$$\sum_{r=1}^n \sum_{\substack{\Rate^{(r)}\in\Lambda\\\|\Rate^{(r)}\|_0=r}} \delta^r \leq 1.$$ 
For every fixed $r$, there are exactly ${n \choose r}$ subspaces of dimension $r$, and each subspace contains exactly $\left(\frac{L_0}{\sqrt{\delta}}\right)^r$ distinct elements of $\Lambda$. Therefore, as long as
\begin{equation}\label{eq:delta_min}\delta\leq \left(2n\,L_0 \right)^{-2},\end{equation} then
 $$\sum_{r=1}^n {n \choose r} \left(L_0\,\sqrt{\delta}\right)^r \leq \sum_{r=0}^n \left(n\,L_0\,\sqrt{\delta}\right)^r \leq \sum_{r=1}^n \frac{1}{2^r}\leq 1,$$ and Kraft's inequality is satisfied.

Using the fact that $k \log(\inDim/k) = O(k d)$, we can bound the minimum over
$\Rate \in \Lambda$ in \eqref{eq:pMLE_bound} from above by
\begin{align*}
&\min_{1 \le r \le \inDim} \left[ k d r^{-2\alpha} + \frac{r\,k\log(\delta^{-1})}{\samp\res }\right] \\
&\qquad = O \left(k\,d^{\frac{1}{2\alpha+1}}\right)\left(\frac{\log(\delta^{-1})}{\samp\res}\right)^{\frac{2\alpha}{2\alpha+1}} \\
&\qquad = O \left(k\,d^{\frac{1}{2\alpha+1}}\right) \left(\frac{\log n}{\samp\res}\right)^{\frac{2\alpha}{2\alpha+1}}
\end{align*}
We can now particularize Theorem~\ref{thm:pMLE} to the
power-law case:

\begin{theorem}\label{thm:compressible_pMLE}
\begin{align*}
  & \sup_{\Rates \in \Sigma_{\alpha,L_0}} \Risk{\hRate^\pMLE_\samp}{\Rates}  \nonumber\\
  & \quad = O(k^{-\alpha})  +
 O \left(k^{\frac{1}{2}}\,d^{\frac{1}{4\alpha+2}}\right) \left(\frac{\log n}{\samp\res}\right)^{\frac{\alpha}{2\alpha+1}},
\end{align*}
where the constants implicit in the $O(\cdot)$ notation depend on $L_0$ and $\alpha$.
\end{theorem}
Note that the risk bound here is slightly worse than the benchmark bound of
Theorem~\ref{thm:direct}. However, it should be borne in mind that this bound is based on Theorem~\ref{thm:expander2}, rather than on the potentially much tighter oracle inequality of Theorem~\ref{thm:expander2_main}, since our goal was to express the risk of the pMLE purely in terms of the $\ell_1$ approximation properties of the power-law class $\Sigma_{\alpha,L_0}$. In general, we will expect the actual risk of the pMLE to be much lower than what the conservative bound of Theorem~\ref{thm:compressible_pMLE} predicts. Indeed, as we will see in Section~\ref{sec:exp}, the pMLE approach obtains higher empirical accuracy than the direct method. But first we show how
the pMLE can be approximated efficiently with proper
preprocessing of the observed counts $\Cts_\samp$ based on the structure
of $\Graph_{k,\inDim}$.

\subsection{Efficient pMLE approximation}
\label{sec:alg}

In this section we present an efficient algorithm for approximating
the pMLE estimate. The algorithm consists of two phases: (1) first,
we preprocess $\Cts_\samp$ to isolate a subset $\Vin_1$ of $\Vin =
\{1,\ldots,\inDim\}$ which is sufficiently small and is guaranteed to
contain the locations of the $k$ largest entries of $\Rates$ (the
whales); (2) then we construct a set $\Lambda$ of candidate estimators
whose support sets lie in $\Vin_1$, together with an appropriate
penalty, and perform pMLE over this reduced set.

The success of this approach hinges on the assumption that the
magnitude of the smallest whale is sufficiently large compared to the magnitude of the largest minnow. Specifically, we make the following
assumption: Let $S \subset \Vin$ contain the locations of the $k$
largest coordinates of $\Rates$. Then we require that
\begin{align}\label{eq:SNR_assumption}
\min_{i \in S} \Rates_i > 9\bound.
\end{align}
Recall that $D=O\left(\frac{nd}{m}\right)=O\left(\frac{n}{k}\right)$ is the right degree of the expander graph. One way to think about \eqref{eq:SNR_assumption} is in terms of a
signal-to-noise ratio, which must be strictly larger than $9\,D$. We also require $\samp\res$ to be sufficiently large, so that
\begin{align}\label{eq:final_assumption}
\frac{\samp\res}{m}\,\bound\geq \frac{\log\left(mn\right)}{2}.
\end{align}
Finally, we perturb our expander a bit as follows:
choose an integer $k' > 0$ so that
\begin{align}\label{eq:expander_perturb}
k' \ge \max \left\{ \frac{16(kd + 1)}{15d}, 2k\right\}.
\end{align}
Then we replace our original $(2k,1/16)$-expander $\Graph_{k,\inDim}$ with left-degree $d$
with a $(k',1/16)$-expander $\Graph'_{k',\inDim}$ with the same left degree.The resulting
procedure, displayed below as Algorithm~\ref{alg1}, has the following
guarantees:

\begin{algorithm}[ht]
\caption{Efficient pMLE approximation algorithm}
\textbf{Input:} Measurement vector $\Cts_\T$, and the sensing matrix
$\AdjMat$.  \textbf{Output:} An approximation $\has$
\begin{algorithmic}
\label{alg1}
\STATE Let $\Vout_1$ consist of the locations of the $kd$ largest
elements of $\Cts_\T$ and let $\Vout_2 = \Vout \backslash \Vout_1$.
\STATE Let $\Vin_2$ contain the set of all variable nodes that have at
least one neighbor in $\Vout_2$ and let $\Vin_1\nobreak= \Vin
\backslash \Vin_2$.  \STATE Construct a candidate set of estimators
$\Lambda$ with support in $\Vin_1$ and a penalty $\pen(\cdot)$ over
$\Lambda$.  \STATE \label{finalstep}Output the pMLE $\hRate$.
\end{algorithmic}
\end{algorithm}

\begin{theorem}
 \label{thm:sina}
Suppose the assumptions~\eqref{eq:SNR_assumption}, \eqref{eq:final_assumption}, and \eqref{eq:expander_perturb} hold. Then with probability at least $1-\frac{1}{n}$ the set $\Vin_1$ constructed by Algorithm~\ref{alg1} has the
 following properties: (1) $S \subset \Vin_1$; (2) $|\Vin_1| \le kd$;
 (3) $\Vin_1$ can be found in time $O(\outDim \log \outDim + \inDim d)$.
  \end{theorem}
  \begin{figure*}[ht]
  \centering \subfigure[$\alpha=1$]{\includegraphics[width=0.30\textwidth]{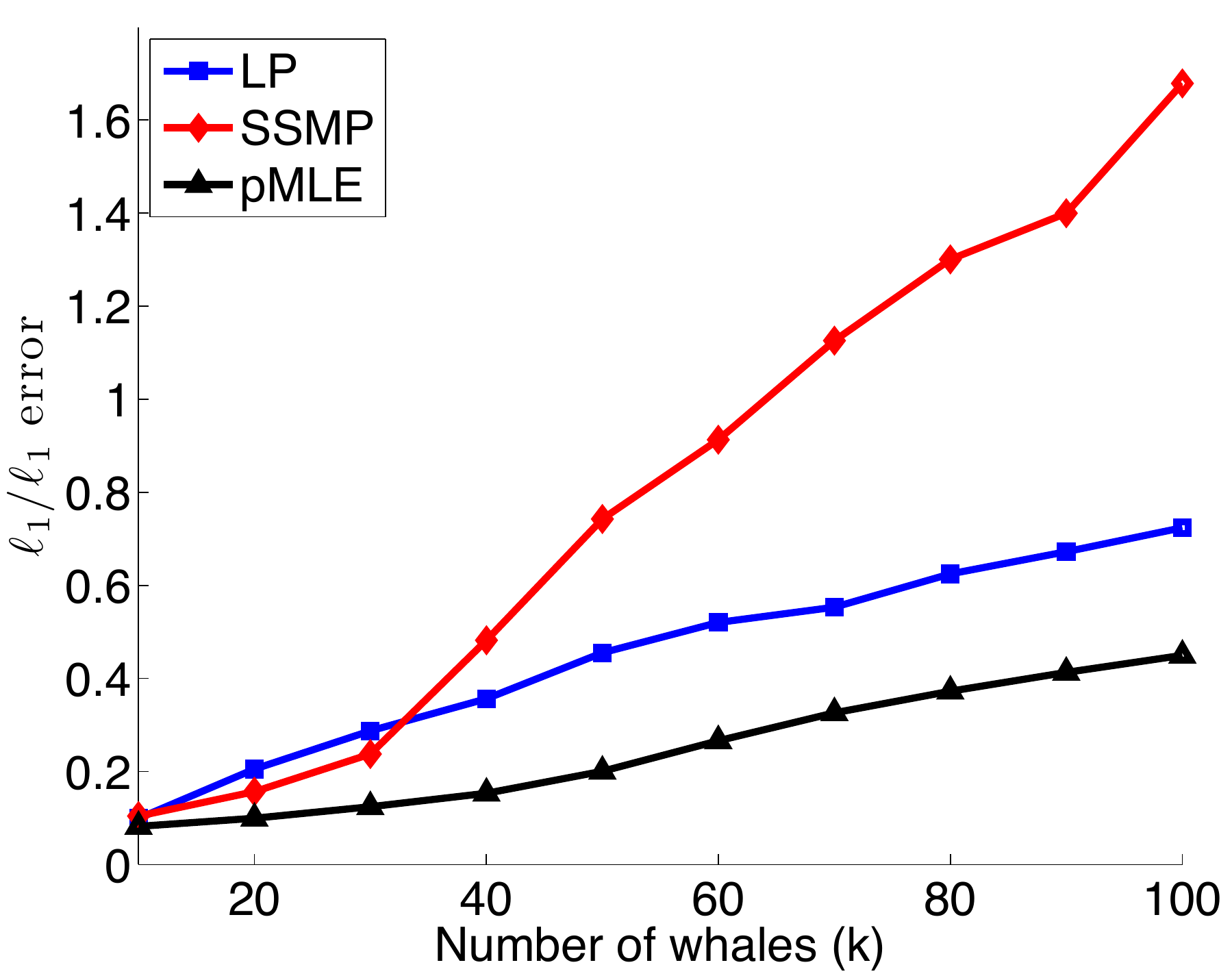}
    \label{fig1}} \subfigure[$\alpha=1.5$]
{\includegraphics[width=0.30\textwidth]{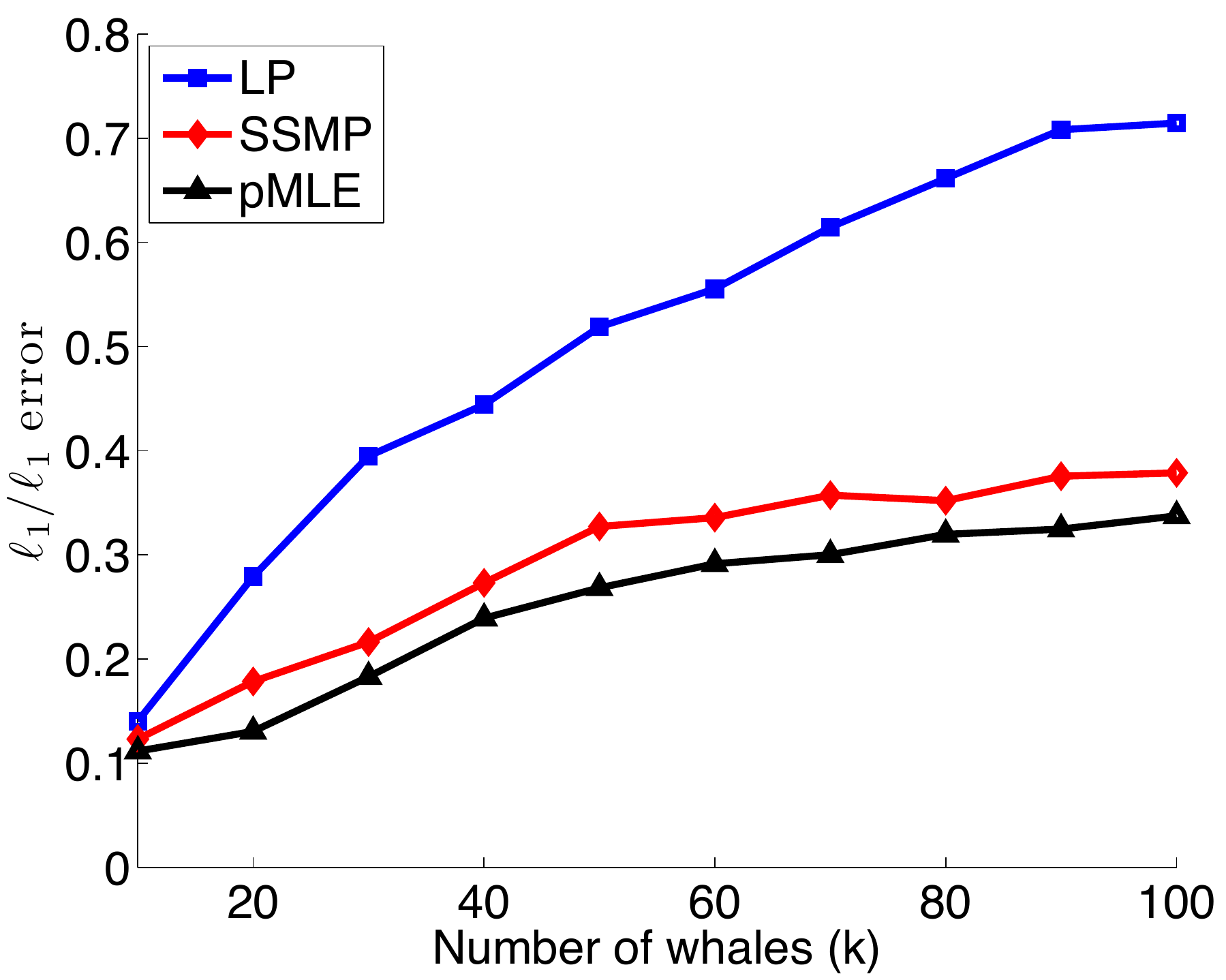}
\label{fig2}}
 \subfigure[$\alpha=2$]{\includegraphics[width=0.30\textwidth]{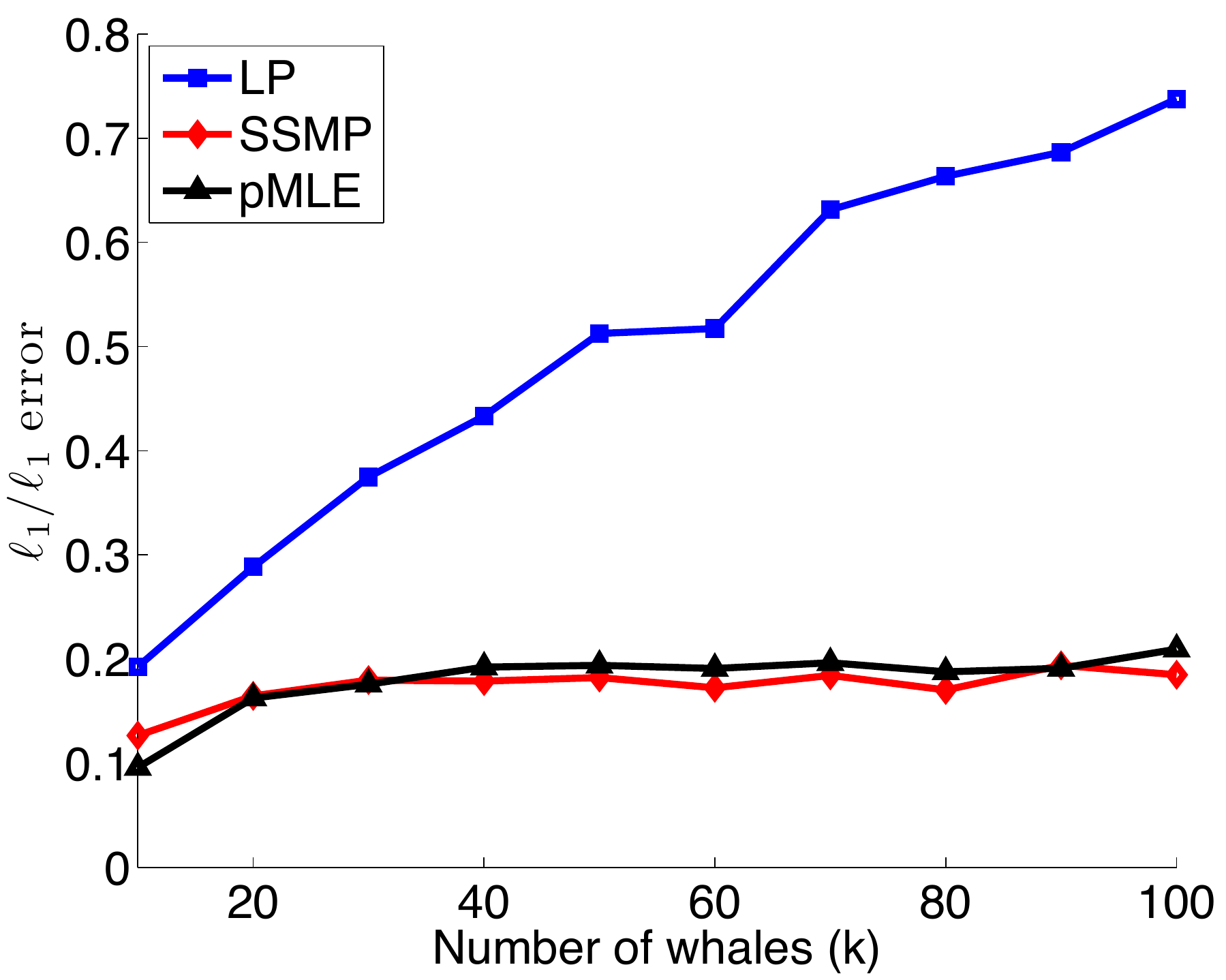}
\label{fig3}}
\caption{Relative $\ell_1$ error as a function of
number of whales $k$, for $\ell_1$-magic (LP), SSMP
  and pMLE for different choices of the power-law exponent $\alpha$. The number of flows $\n=5000$, the number of counters
  $\m=800$, and the number of updates is $40$.}
\label{fig1-3}
\end{figure*}

 \begin{figure*}[ht]
  \centering \subfigure[$\alpha=1$]{\includegraphics[width=0.30\textwidth]{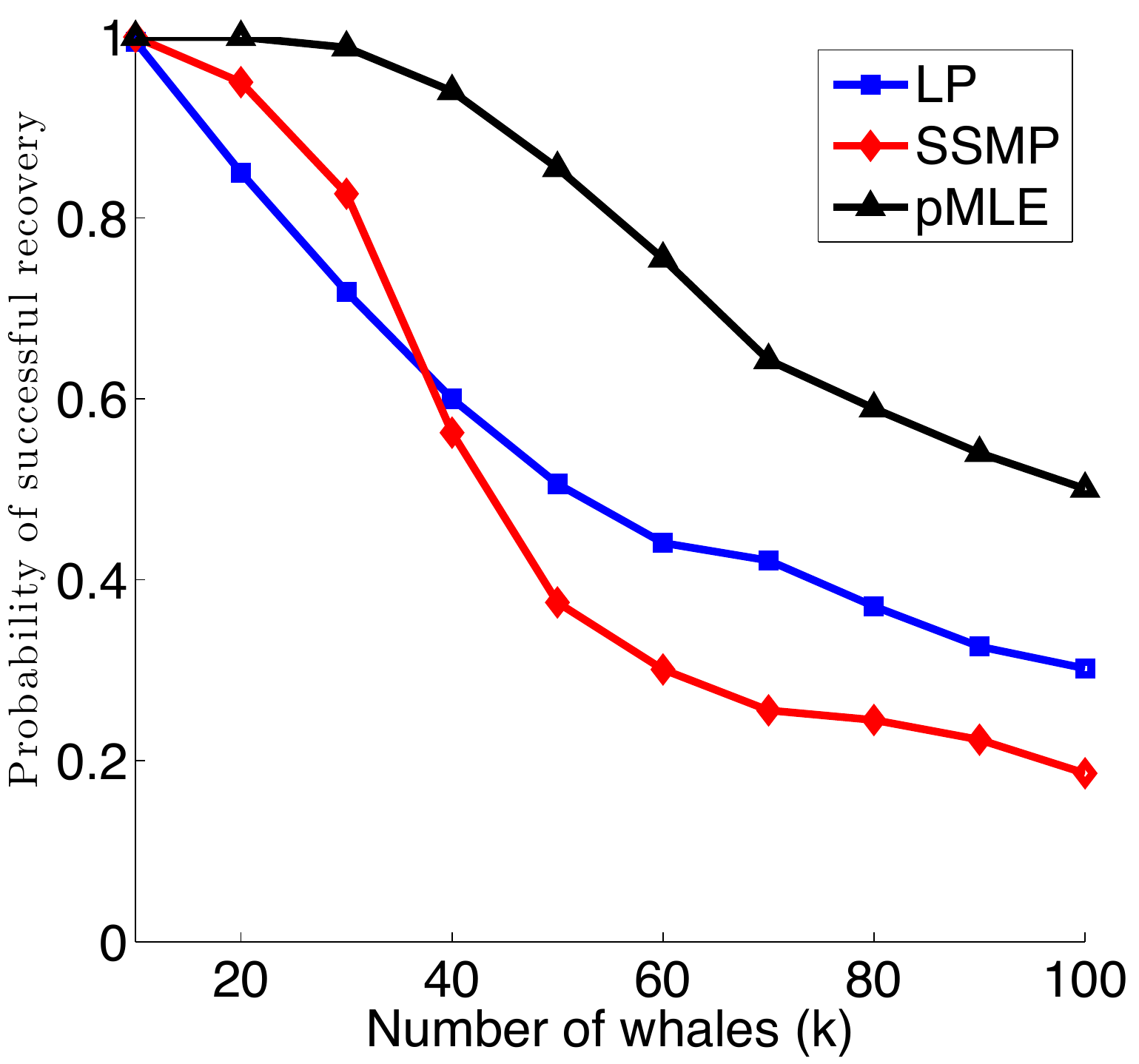}
    \label{fig4}} \subfigure[$\alpha=1.5$]
{\includegraphics[width=0.30\textwidth]{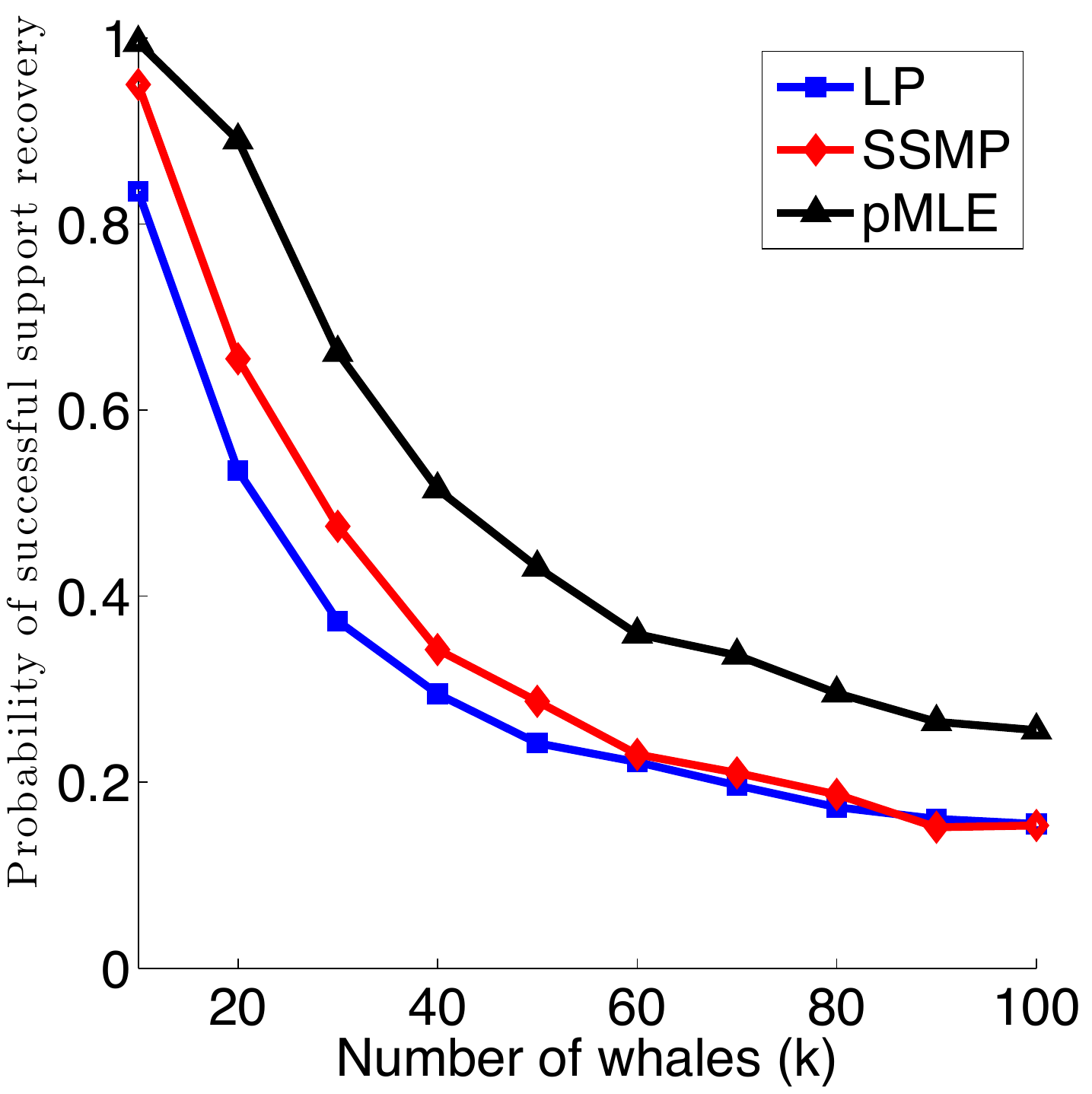}
\label{fig5}}
 \subfigure[$\alpha=2$]{\includegraphics[width=0.30\textwidth]{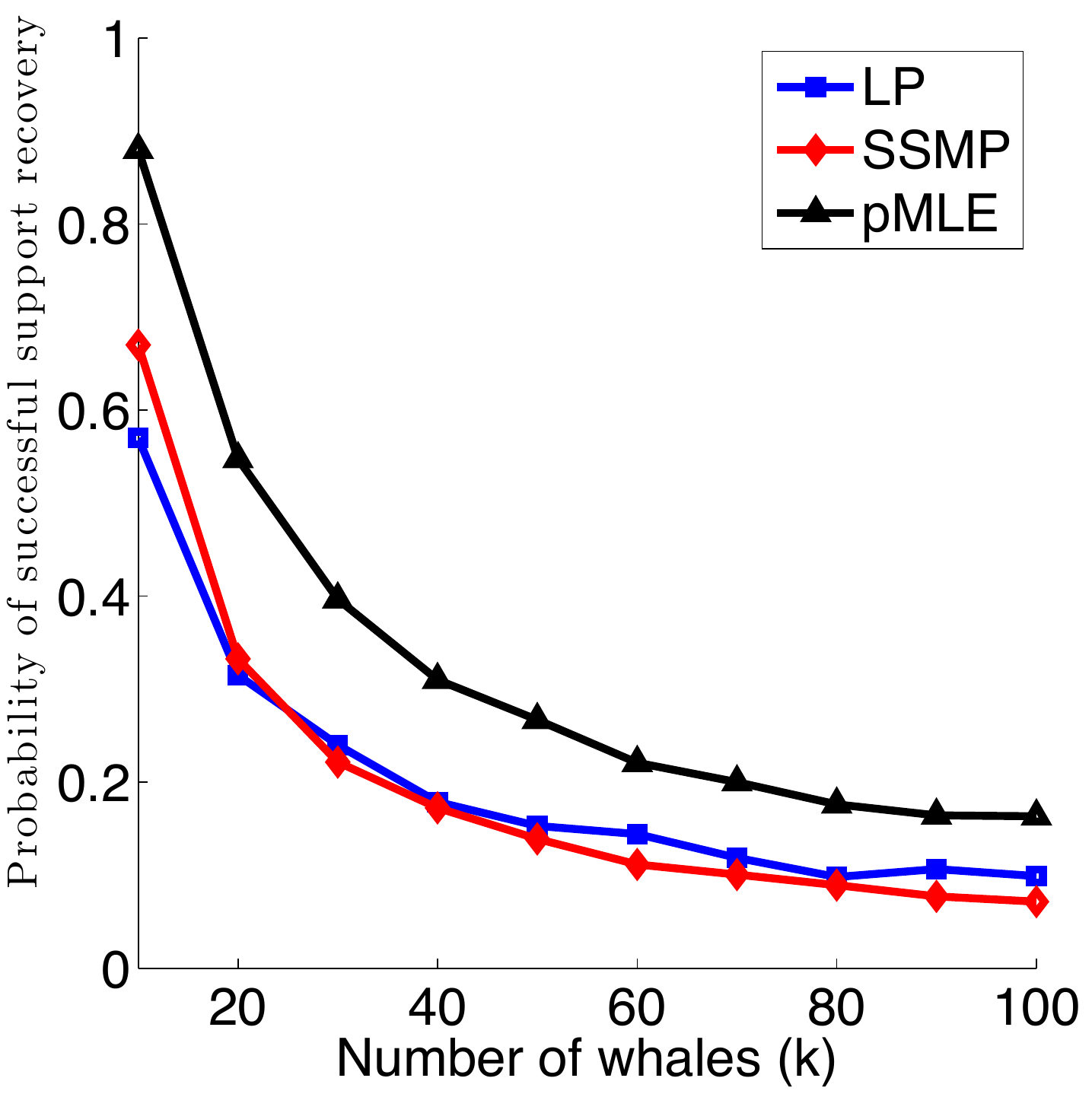}
\label{fig6}}
\caption{Probability of successful support recovery as
  a function of number of whales
  $k$, for $\ell_1$-magic (LP), SSMP
  and pMLE for different choices of the power-law exponent $\alpha$. The number of flows $\n=5000$, the number of counters
  $\m=800$, and the number of updates is $40$.}
\label{fig4-6}
\end{figure*}
 \begin{IEEEproof}
   (1)
    First fix a measurement node $j\in\Vout$. Recall that $\Cts_{\samp,j}$ is a Poisson random variable with mean $\frac{\samp\res}{m} \left(\AdjMat \Rates\right)_j$. By the same argument as in Remark~\ref{rem:L1_estimation},  $\sqrt{\Cts_{\samp,j}}$ is approximately normally distributed with mean $\approx\sqrt{\frac{\samp\res}{m} \left(\AdjMat \Rates\right)_j}$, and with variance $\approx \frac{1}{4}$. Hence, it follows from Mill's inequality and the union bound that for every positive $t$ 
    \begin{equation*}
    \Pr\left[\exists j: \left|\sqrt{\Cts_{\samp,j}}-\sqrt{\frac{\samp\res}{m}  \left(\AdjMat \Rates\right)_j}\right|> t \right]\lesssim \frac{ m e^{-2t^2}}{\sqrt{2\pi} t}.
    \end{equation*}
    If $j$ is a neighbor of $S$, then $\left(\AdjMat \Rates\right)_j\geq \min_{i \in S} \Rates_i$; whereas if $j$ is not connected to $S$, then $\left(\AdjMat \Rates\right)_j\leq \bound$. Hence, by setting $t =\sqrt{\frac{\log\left(mn\right)}{2}}$ (where w.l.o.g.~we assume that $t \ge 1$), we conclude that, with probability at least $1-\frac{1}{n}$, for every measurement node $j$ the following holds:
    \begin{itemize}
    \item If $j$ is a neighbor of $S$, then $$\sqrt{\Cts_{\samp,j}}\geq\sqrt{\frac{\samp\res}{m}  \min_{i \in S} \Rates_i}-\sqrt{\frac{\log\left(mn\right)}{2}}.$$
    \item If $j$ is not connected to $S$, then $$\sqrt{\Cts_{\samp,j}}\leq\sqrt{\frac{\samp\res}{m}  \bound}+\sqrt{\frac{\log\left(mn\right)}{2}}.$$
    \end{itemize}
    Consequently, by virtue of \eqref{eq:SNR_assumption} and \eqref{eq:final_assumption}, with probability at least $1-\frac{1}{n}$ every element of $\Cts_\T$ that is a neighbor of $S$ has larger magnitude than every element of $\Cts_\T$ that is not a neighbor of $S$. 

(2) Suppose, to the contrary, that $|\Vin_1| > kd$. Let $\Vin'_1
\subseteq \Vin_1$ be any subset of size $kd+1$. Now, Lemma~3.6 in
\cite{Khaj} states that, provided $\eps \le 1-1/d$, then every
$(\ell,\eps)$-expander with left degree $d$ is also a
$(\ell(1-\eps)d,1-1/d)$-expander with left degree $d$. We apply this
result to our $(k',1/16)$-expander, where $k'$ satisfies
\eqref{eq:expander_perturb}, to see that it is also a
$(kd+1,1-1/d)$-expander. Therefore, for the set $\Vin'_1$ we must have
$|\cN(\Vin'_1)| \ge |\Vin'_1| = kd+1$. On the other hand,
$\cN(\Vin'_1) \subset \Vout_1$, so $|\cN(\Vin'_1)| \le kd$. This is a
contradiction, hence we must have $|\Vin_1| \le kd$.

(3) Finding the sets $\Vout_1$ and $\Vout_2$ can be done in $O(\outDim
\log \outDim)$ time by sorting $\Cts_\T$. The set $\Vin_1$ can then
can be found in time $O(\inDim d)$, by sequentially eliminating all
nodes connected to each node in $\Vout_2$.
 \end{IEEEproof}
  
Having identified the set $\Vin_1$, we can reduce the pMLE
optimization only to those candidates whose support sets lie in
$\Vin_1$. More precisely, if we originally start with a sufficiently
rich class of estimators $\tilde{\Lambda}$, then the new feasible set
can be reduced to
 $$\Lambda \deq \left\{ \Rate\in\tilde{\Lambda} :  {\rm Supp}(\Rate)\subset \Vin_1  \right\}.$$ 
 Hence, by extracting the set $\Vin_1$, we can significantly reduce
 the complexity of finding the pMLE estimate. If $|\Lambda|$ is small,
 the optimization can be performed by brute-force search in
 $O(|\Lambda|)$ time. Otherwise, since $|\Vin_1| \le kd$, we can use
 the quantization technique from the preceding section with quantizer
 resolution $\sqrt{\delta}$ to construct a $\Lambda$ of size at most
 $(L_0/\sqrt{\delta})^{kd}$. In this case, we can even assign the
 uniform penalty
 $$
 \pen(\Rate) = \log |\Lambda| = O\left( k \log(\inDim/k)
   \log(\delta^{-1})\right),
 $$
which amounts to a vanilla MLE over $\Lambda$.
 
\subsection{Empirical performance}
\label{sec:exp} 
Here we compare penalized MLE with $\ell_1$-magic \cite{l1}, a
universal $\ell_1$ minimization method, and with SSMP \cite{ssmp}, an
alternative method that employs combinatorial
optimization. $\ell_1$-magic and SSMP both compute the ``direct''
estimator. The pMLE
estimate is computed using Algorithm~\ref{alg1} above. For the ease of computation, the candidate set $\Lambda$ is approximated by the convex set of all positive vectors with bounded $\ell_1$ norm, and the CVX package \cite{cvx1,cvx2} is used to directly solve the pMLE objective function with $\pen(\theta)=\|\theta\|_1$.

\begin{figure}[ht]
  \centering \subfigure[Relative $\ell_1$ error as a function of
number of updates $\nu$.]{\includegraphics[width=0.45\textwidth]{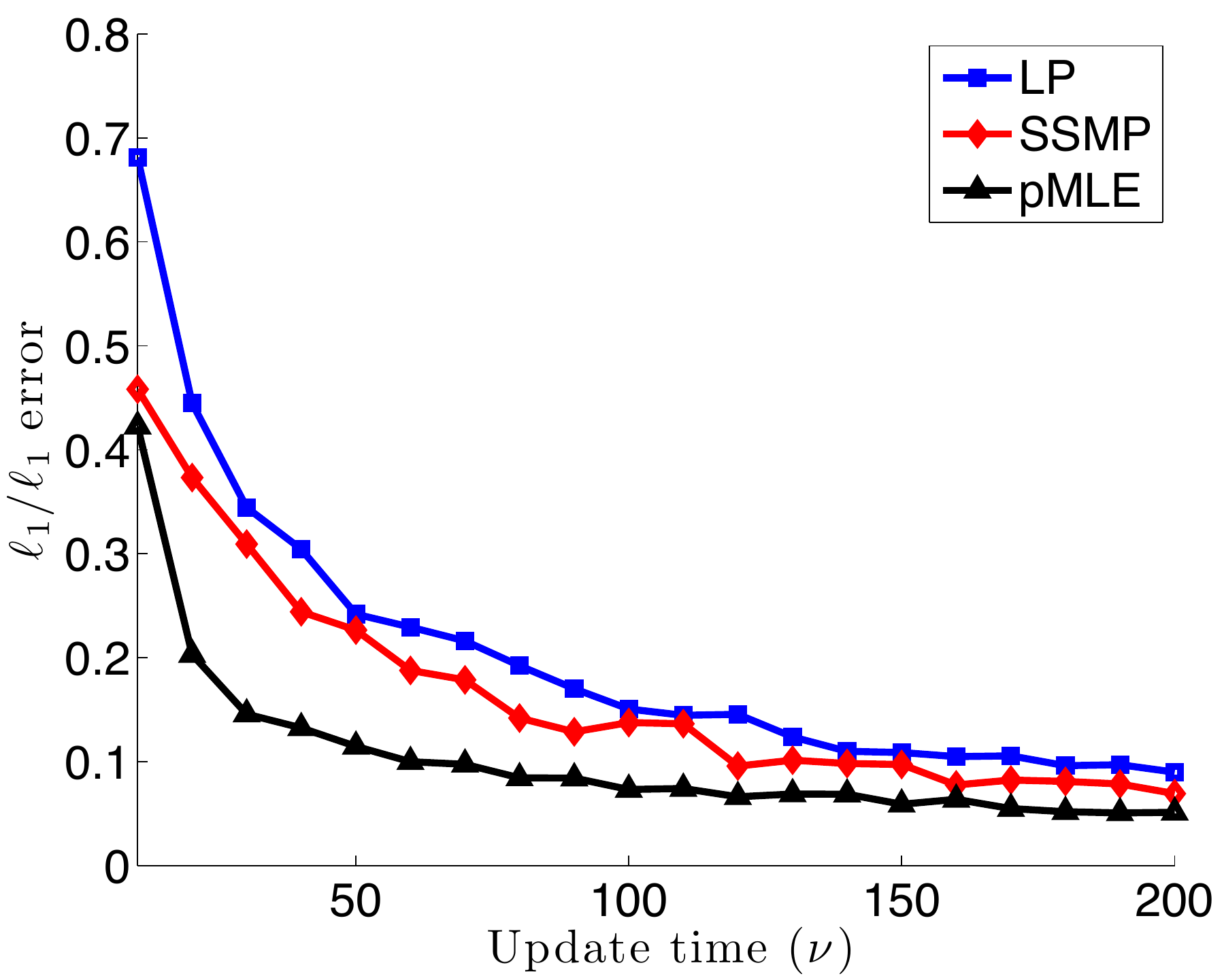}
    \label{fig7}} 
    \subfigure[Probability of successful support recovery as
  a function of number of updates $\nu$.]
{\includegraphics[width=0.45\textwidth]{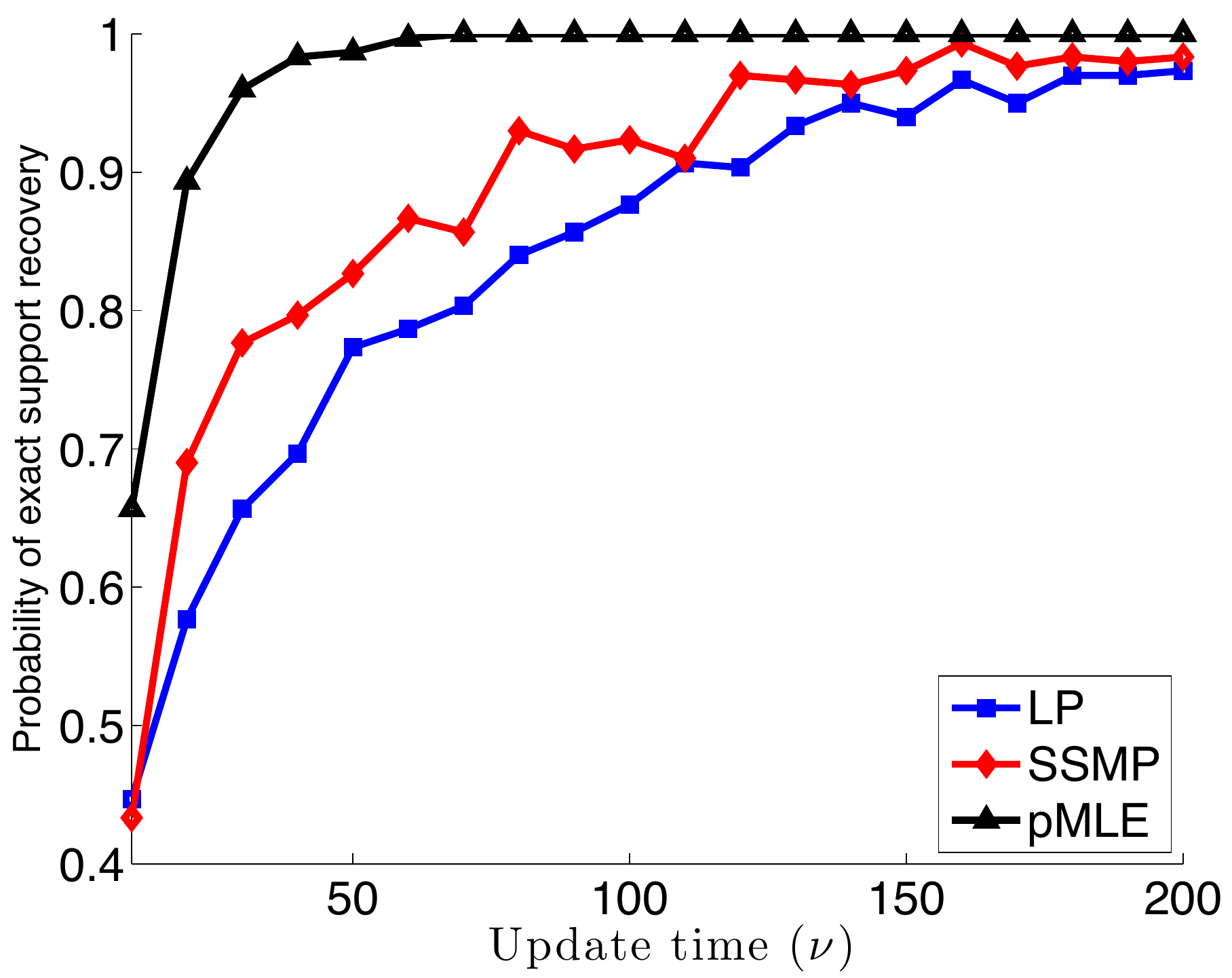}
\label{fig8}}
\caption{Performance of $\ell_1$-magic, SSMP
  and pMLE algorithms as a function of the number of updates $\nu$. The number of flows $\n=5000$, the number of counters
  $\m=800$, and the number of whales is $k=30$. There are $k$ whales
 whose magnitudes are assigned according to a power-law distribution with $\alpha=1$, and the remaining entries are minnows with
  magnitudes determined by a ${\cal N}(0, 10^{-6})$ random variable.}
\label{fig7-8}
\end{figure}

Figures~\ref{fig1} through~\ref{fig8} report the results of numerical
experiments, where the goal is to identify the $k$ largest entries in
the rate vector from the measured data. Since a random graph is, with
overwhelming probability, an expander graph, each experiment was
repeated $30$ times using independent sparse random graphs with $d=8$.
  
We also used the following process to generate the rate vector. First, given the power-law exponent $\alpha$, the magnitudes of the $k$ whales were chosen according to a power-law distribution with parameter $\alpha$. The positions of the $k$ whales were then chosen uniformly at random. Finally the $n-k$ minnows were sampled independently from a ${\cal N}(0,10^{-6})$ distribution (negative samples were replaced by their absolute values). Thus, given the locations of the $k$ whales, their magnitudes decay according to a {\em truncated} power law (with the cut-off at $k$), while the magnitudes of the minnows represent a noisy background.
Figure~\ref{fig1-3} shows the relative $\ell_1$ error ($\|\Rate-\hRate_\samp\|_1/\|\Rate\|_1$) of the three above algorithms as a function of $k$. Note that in all cases $\alpha=1$, $\alpha=1.5$, and $\alpha=2$, the pMLE algorithm provides lower $\ell_1$ errors. Similarly, Figure~\ref{fig4-6} reports the probability of exact recovery as a function of $k$. Again, it turns out that in all three cases the pMLE algorithm has higher probability of exact support recovery compared to the two direct algorithms.

We also analyzed the impact of changing the number of updates on the accuracy of the three above algorithms. The results are demonstrated in Figure~\ref{fig7-8}. Here we fixed the number of whales to $k=30$, and changed the number of updates from $10$ to $200$. It turned out that as the number of updates $\nu$ increases, the relative $\ell_1$ errors of all three algorithms decrease and their probability of exact support recovery consistently increase. Moreover, the pMLE algorithm always outperforms the $\ell_1$-magic (LP), and SSMP algorithms.

\section{Conclusions}

In this paper we investigated  expander-based sensing as an
alternative to dense random sensing in the presence of Poisson noise.  Even
though the Poisson model is essential in some applications, it
presents several challenges as the noise is not bounded, or even
as concentrated as Gaussian noise, and is signal-dependent.  Here we
proposed using normalized adjacency matrices of expander graphs as an
alternative construction of sensing matrices, and we showed that the
binary nature and the RIP-1 property of these matrices yield provable
consistency for a MAP reconstruction algorithm.

The compressed sensing algorithms based on Poisson observations and
expander-graph sensing matrices provide a useful mechanism for
accurately and robustly estimating a collection of flow rates with
relatively few counters. These techniques have the potential to
significantly reduce the cost of hardware required for flow rate
estimation. While previous approaches assumed packet counts matched
the flow rates exactly or that flow rates were i.i.d., the approach in
this paper accounts for the Poisson nature of packet counts with
relatively mild assumptions about the underlying flow rates
(i.e.,~that only a small fraction of them are large).

The ``direct'' estimation method (in which first the vector of flow
counts is estimated using a linear program, and then the underlying
flow rates are estimated using Poisson maximum likelihood) is
juxtaposed with an ``indirect'' method (in which the flow rates are
estimated in one pass from the compressive Poisson measurements using
penalized likelihood estimation). 

The methods in this paper, along with related results in this area,
are designed for settings in which the flow rates are sufficiently
stationary, so that they can be accurately estimated in a fixed time
window. Future directions include extending these approaches to a more
realistic setting in which the flow rates evolve over time. In this
case, the time window over which packets should be counted may be
relatively short, but this can be mitigated by exploiting estimates of
the flow rates in earlier time windows.

\appendices
\section{Observation models in Poisson inverse problems}
\label{app:shot}

In \eqref{eq:obs} and all the subsequent analysis in this paper,
  we assume 
\begin{align*}
y \sim \text{Poisson}(\Phi \ts).
\end{align*}
However, one might question how accurately this models the physical
systems of interest, such as a photon-limited imaging system or a
router. In particular, we may prefer to think of only a small number
of events (e.g., photons or packets) being incident upon our system,
and the system then rerouting those events to a detector. In this
appendix, we compare the statistical properties of these two
models. Let $z_{j,i}$ denote the number of events traveling from
location $i$ in the source ($\ts$) to location $j$ on the
detector. Also, in this appendix let us assume $\Phi$ is a stochastic
matrix, i.e., each column of $\Phi$ sums to one; in general, most
elements of $\Phi$ are going to be less than one. Physically, this
assumption means that every event incident on the system hits some
element of the detector array. Armed with these assumptions, we can
think of $\Phi_{j,i}$ as the probability of events from location $i$
in $\ts$ being transmitted to location $j$ in the observation vector
$y$.

We consider two observation models:
\begin{align*}
\mbox{Model A:}&& z_{j,i} &\sim \text{Poisson}(\Phi_{j,i} \ts_i) \\
&& y_j & \deq \sum_{i=1}^n z_{j,i}\\
\mbox{Model B:}&& w &\sim \text{Poisson}(\ts) \\
&& \{ z_{j,i} \}_{i=1}^n &\sim \text{Multinomial}(w_i,\{\Phi_{j,i}\}_{i=1}^n)\\
&& y_j & \deq \sum_{i=1}^n z_{j,i},
\end{align*}
where in both models all the components $z_{j,i}$ of $z$ are mutually conditionally independent given the appropriate parameters. Model A roughly corresponds to the model we consider throughout the
paper; Model B corresponds to considering Poisson realizations with
intensity $\ts$ (denoted $w$) incident upon our system and then
redirected to different detector elements via $\Phi$. We model this redirection process  with a multinomial distribution. While the model $y \sim
\text{Poisson}(\Phi\ts)$ is slightly different from Model A, the
following analysis will provide valuable insight into discrete event
counting systems.

We now show that the distribution of $z$ is the same in Models A
  and B. First note that
\begin{align}
y_j &\equiv \sum_{i=1}^n z_{j,i} &\mbox{and}&& w_i &\equiv \sum_{j=1}^m z_{j,i} \label{eq:z}.
\end{align}
Under Model A, we have
\begin{align}
p(z|\ts) &= \prod_{i=1}^n \prod_{j=1}^m \frac{e^{-\Phi_{j,i}\ts_i}
  (\Phi_{j,i}\ts_i)^{z_{j,i}}}{z_{j,i}!} \nonumber \\
&= \prod_{i=1}^n \left(\prod_{j=1}^m
  \frac{\Phi_{j,i}^{z_{j,i}}}{z_{j,i}!} \right) e^{-\sum_{j=1}^m
  \Phi_{j,i} \ts_i} (\ts_i)^{\sum_{j=1}^m z_{j,i}}\nonumber \\
&=\prod_{i=1}^n \left(\prod_{j=1}^m
  \frac{\Phi_{j,i}^{z_{j,i}}}{z_{j,i}!} \right) e^{-\ts_i} (\ts_i)^{w_i} \label{eq:ModelA}
\end{align}
where in the last step we used \eqref{eq:z} and the assumption that
$\sum_{j=1}^m \Phi_{j,i} = 1$.
Under Model B, we have 
\begin{align}
p(z|w) &= \left\{\begin{array}{ll}
\displaystyle \prod_{i=1}^n w_i! \prod_{j=1}^m \frac{\Phi_{j,i}^{z_{j,i}}}{z_{j,i}!},
& \text{if } \displaystyle \sum_{j=1}^m z_{j,i} = w_i \, \forall i \\
0, & \text{otherwise}
\end{array} \right.
\nonumber \\
p(w|\ts) &= \prod_{i=1}^n \frac{e^{-\ts_i} (\ts_i)^{w_i}}{w_i!} \nonumber \\
p(z|\ts) &= \sum_{v \in \Zplus^m: \sum_j z_{j,i} = v_i} p(z|v) p(v|\ts) \nonumber
\\
&= \prod_{i=1}^n \prod_{j=1}^m w_i!  \frac{\Phi_{j,i}^{z_{j,i}}}{z_{j,i}!}
\frac{e^{-\ts_i} (\ts_i)^{w_i}}{w_i!} \nonumber \\
&= \prod_{i=1}^n \left(\prod_{j=1}^m \frac{\Phi_{j,i}^{z_{j,i}}}{z_{j,i}!}\right)
e^{-\ts_i} (\ts_i)^{w_i}. \label{eq:ModelC}
\end{align}
The fourth line uses \eqref{eq:z}.  Since \eqref{eq:ModelA} and
\eqref{eq:ModelC} are the same, we have shown that Models A and B are
statistically equivalent. While Model B may be more intuitively
appealing based on our physical understanding of how these systems
operate, using Model A for our analysis and algorithm development is
just as accurate and mathematically more direct.

\section{Proof of Proposition~\ref{piotr}}
\label{app:piotr}

Let $y=u-v$, let $S \subset \{1,\ldots,\inDim\}$ denote the positions of the $k$ largest (in magnitude) coordinates of $y$, and enumerate the complementary set $S^c$ as $i_1,i_2,\ldots,i_{\inDim -k}$ in decreasing order of magnitude of $|y_{i_j}|, j = 1,\ldots,\inDim -k $. Let us partition the set $S^c$ into adjacent blocks $S_1,\ldots,S_t$, such that all blocks (but possibly $S_t$) have size $k$. Also let $S_0 = S$. Let $\At$ be a submatrix of $\AdjMat$ containing
  rows from ${\cal N}(S)$. Then, following the argument of Berinde et al.~\cite{newIndyk}, which also goes back to Sipser and Spielman \cite{SS}, we have the following chain of inequalities:
\begin{align*}
& \|\AdjMat y\|_1 \geq \|\At y\|_1\\ 
\nonumber &\quad \geq \|\At y_S\|_1 -\sum_{i=1}^t \sum_{(j,l) \in \Edges: j \in S_i,\,l\in {\cal N}(S)}|y_j| \\ \nonumber 
&\quad \geq d(1-2\epsilon) \|y_S\|_1 -\sum_{i=1}^t \sum_{(j,l) \in \Edges: j \in S_i,\,l\in {\cal N}(S)}\frac{\|y_{S_{i-1}}\|_1}{k}
\\ \nonumber
&\quad \geq d(1-2\epsilon) \|y_S\|_1-2kd\epsilon \sum_{i=1}^t \frac{\|y_{S_{i-1}}\|_1}{k}
\\ \nonumber 
&\quad \geq d(1-2\epsilon) \|y_S\|_1-2d\epsilon \|y\|_1.
\end{align*}
Most of the steps are straightforward consequences of the definitions, the triangle inequality, or the RIP-$1$ property. The fourth inequality follows from the following fact. Since we are dealing with a $(2k,\epsilon)$-expander and since $|S \cup S_i| \le 2k$ for every $i=0,\ldots,t$, we must have $|{\cal N}(S \cup S_i)| \ge d(1-\epsilon)|S \cup S_i|$. Therefore, at most $2kd\epsilon$ edges can cross from each $S_i$ to ${\cal N}(S)$. From the above estimate, we obtain
\begin{equation}
\label{lemma}
\|\AdjMat u- \AdjMat v\|_1+2d\epsilon\|y\|_1 \geq (1-2\epsilon)d \|y_S\|_1.
\end{equation}
Using the assumption that $\| u \|_1 \ge \| v \|_1 - \Delta$, the triangle inequality, and the fact that $\| u_{S^c} \|_1 = \kerr{u}$, we obtain
\begin{align*}
	\| u \|_1 &\ge \| v \|_1 - \Delta \\
	&= \| u - y \|_1 - \Delta \\
	&= \| (u-y)_S\|_1 + \|(u-y)_{S^c} \|_1 - \Delta \\
	&\ge \| u_S \|_1 - \| y_S \|_1 + \| u_{S^c} \|_1 - \| y_{S^c} \|_1 - \Delta \\
	&= \| u \|_1 - 2 \| u_{S^c} \|_1 + \| y \|_1 - 2 \| y_{S} \|_1 - \Delta \\
	&= \| u \|_1 - 2\kerr{u} + \| y \|_1 - 2 \| y_S \|_1 - \Delta,
\end{align*}
which yields
\begin{align*}
	\| y \|_1 \le 2\kerr{u} + 2 \| y_S \|_1 + \Delta.
\end{align*}
Using \eqref{lemma} to bound $\| y_S \|_1$, we further obtain
\begin{align*}
	\| y \|_1 \le 2\kerr{u} + \frac{2\| \AdjMat u - \AdjMat v \|_1 + 4d\epsilon \| y \|_1}{(1-2\epsilon)d} + \Delta.
\end{align*}
Rearranging this inequality completes the proof.

\section{Technical lemmas}
\label{app:lemmas}

\begin{lemma}
\label{l1}
Any $\tb \in \Theta_L$ satisfies the bound
\begin{align*}
\|\nAdjMat(\ts-\tb)\|_1^2 \leq 4L \sum_{i=1}^m \left|(\nAdjMat\ts)_i^{1/2}-(\nAdjMat\tb)_i^{1/2} \right|^2.
\end{align*}
\end{lemma}
\begin{IEEEproof}
From Lemma~\ref{RIP} it follows that
\begin{align}\label{eq:mona_1}
	\| \Phi \tb \|_1 \le \| \tb \|_1 \le L, \qquad \forall \tb \in \Theta_L.
\end{align}
	Let $\ybs\deq\nAdjMat \ts$ and $\beta\deq\nAdjMat \tb$. Then
\begin{align*}
 \|\ybs-\beta\|_1^2 &= \left(\sum_{i=1}^m \left|\ybs_i-\beta_i\right|\right)^2\\ \nonumber
 &= \left(\sum_{i=1}^m \left|\ybs^{\half}_i-\beta_i^{\half} \right|. \left| \ybs_i^{\half}+\ybh_i^{\half} \right|\right)^2
\\ \nonumber &\leq \sum_{i,j=1}^m \left|\ybs_i^{\half}-\beta_i^{\half} \right|^2. \left| \ybs_j^{\half}+\beta_j^{\half} \right|^2
\\ \nonumber &\leq
2\sum_{i=1}^m\left| \ybs_i^{\half}-\beta_i^{\half} \right|^2. \sum_{j=1}^m \left| \ybs_j+\beta_j \right|
\\ \nonumber &=
2\sum_{i=1}^m \left|\ybs_i^{\half}-\beta_i^{\half} \right|^2. \left(\|\ybs\|_1+\|\beta\|_1 \right)
\\ \nonumber &\leq
4L  \sum_{i=1}^m \left|\ybs_i^{\half}-\beta_i^{\half} \right|^2.
\\ \nonumber &=
4L \sum_{i=1}^m \left|(\A f^*)_i^{\half}-(\A f)_i^{\half} \right|^2.
\end{align*}
The first and the second inequalities are by Cauchy--Schwarz, while the third inequality is a consequence of Eq.~(\ref{eq:mona_1}).\end{IEEEproof}

\begin{lemma}
\label{l2}
Let $\that$ be a minimizer in
Eq.~\eqref{eq:PMLE_main}. Then
\begin{eqnarray}
\label{18}
\lefteqn{\Ex_{\nAdjMat\ts}\left[ \sum_{i=1}^m
    \left|(\nAdjMat\ts)_i^{\half}-(\nAdjMat\that)_i^{\half} \right|^2 \right]}\nonumber \\
  &\leq& \min_{\tb \in \Theta_L} \left[
    \RE(\Prob_{\nAdjMat\ts}\parallel
      \Prob_{\nAdjMat\tb})+2\, \pen(\tb)\right].
\end{eqnarray}
\end{lemma}
\begin{IEEEproof}
Using Lemma \ref{int} below with $\la = \Phi \ts$ and $\lb = \Phi \that$ we have
	\begin{eqnarray*}
	\lefteqn{\Ex_{\A\ts}\left[ \sum_{i=1}^m
	  \left|(\A\ts)_i^{\half}-(\A\that)_i^{\half} \right|^2 \right]}
	\\ \nonumber
	&=&\Ex_{\A\ts}\left[ 2\log
	  \frac{1}{\int\sqrt{\Prob_{\A\ts}(\yb)\Prob_{\A\that}(\yb)}d\nu(\yb)}\right]. 
	\end{eqnarray*}
	Clearly
	\begin{align*}
	\int \sqrt{\Prob_{\A\ts}(\yb)\Prob_{\A\that}(\yb)}d\nu(\yb) =
	\Ex_{\A\ts}\left[\sqrt{\frac{\Prob_{\Phii\that}(\yb)}{\Prob_{\Phi\ts}(y)}}\right].
	\end{align*}
	We now provide a bound for this expectation. Let $\tilde{\theta}$ be a minimizer of
	$\RE ( \Prob_{\Phi \ts} \| \Prob_{\Phi \tb}) + 2\, \pen(\tb)$ over $\tb \in \Theta_L$. Then, by definition of $\that$, we have
	\begin{align*}
		\sqrt{\Prob_{\Phi \that}(y)}e^{-\pen(\that)} \ge \sqrt{\Prob_{\Phi \tilde{\theta}}(y)} e^{-\pen(\tilde{\theta})}
	\end{align*}
for every $y$. Consequently,
	\begin{align*}
	\frac{1}{\Ex_{\A\ts}\left[\sqrt{\frac{\Prob_{\Phii\that}(\yb)}{\Prob_{\Phii\ts}(y)}}\right]} \le \frac
	{\sqrt{\Prob_{\A\that}(y)}e^{-\pen(\that)}}
	{\sqrt{\Prob_{\A\tilde{\theta}}(y)}e^{-\pen(\tilde{\theta})} \Ex_{\A\ts}\left[\sqrt{\frac{\Prob_{\Phii\that}(\yb)}{\Prob_{\Phii\ts}(y)}}\right]}, 
	\end{align*}
 We can split the quantity
	$$ 2\Ex_{\Phi\ts}\left[\log\left(\frac{\sqrt{\Prob_{\A\that}(y)}e^{-\pen(\that)}}{\sqrt{\Prob_{\A\tilde{\theta}}(y)}e^{-\pen(\tilde{\theta})}\Ex_{\A\ts}\left[\sqrt{\frac{\Prob_{\Phii\that}(\yb)}{\Prob_{\Phii\ts}(y)}}\right]}\right)\right] 
	$$
	into three terms:
	\begin{align*}
	&\Ex_{\A\ts}
	\left[
	\log
	\left(
		\frac{\Prob_{\A\ts}(y)}
		{\Prob_{\A\tilde{\theta}}(y)}
	\right)
	\right] 
	+ 2\,\pen(\tilde{\theta}) \\
	& + 2\Ex
	\left[
	\log
	\left(
	\frac
	{\sqrt{\Prob_{\A\that}(y)}e^{-\pen(\that)}}
	{\sqrt{\Prob_{\A\ts}(y)}\Ex_{\A\ts}\left[\sqrt{\frac{\Prob_{\Phii\that}(\yb)}{\Prob_{\Phii\ts}(y)}}\right]}\right)\right]
	\end{align*}
	We show that the third term is always nonpositive, which completes
	the proof. Using Jensen's inequality,
	\begin{align*}
	&\Ex
	\left[
	\log
	\left(
	\frac
	{\sqrt{\Prob_{\A\that}(y)}e^{-\pen(\that)}}
	{\sqrt{\Prob_{\A\ts}(y)}\Ex_{\A\ts}\left[\sqrt{\frac{\Prob_{\Phii\that}(\yb)}{\Prob_{\Phii\ts}(y)}}\right]}\right)\right] \\
	& \le \log \left(\Ex
	\left[
	\frac
	{\sqrt{\Prob_{\A\that}(y)}e^{-\pen(\that)}}
	{\sqrt{\Prob_{\A\ts}(y)}\Ex_{\A\ts}\left[\sqrt{\frac{\Prob_{\Phii\that}(\yb)}{\Prob_{\Phii\ts}(y)}}\right]}\right]\right).
	\end{align*}
	Now
	\begin{align*}
	\Ex
	\left[
	\frac
	{\sqrt{\Prob_{\A\that}(y)}e^{-\pen(\that)}}
	{\sqrt{\Prob_{\A\ts}(y)}\Ex_{\A\ts}\left[\sqrt{\frac{\Prob_{\Phii\that}(\yb)}{\Prob_{\Phii\ts}(y)}}\right]}\right] \le \sum_{\tb \in \Theta_L} e^{-\pen(\tb)} \le 1.
	\end{align*}
	Since $\Ex_{\A\ts}
	\left[
	\log
	\left(
		\frac{\Prob_{\A\ts}(y)}
		{\Prob_{\A\tilde{\theta}}(y)}
	\right)
	\right] = \RE(\Prob_{\A \ts} \| \Prob_{\A \tilde{\theta}} )$, we obtain
	\begin{align*}
	& \Ex_{\A\ts}\left[ \sum_{i=1}^m
	  \left|(\A\ts)_i^{\half}-(\A\that)_i^{\half} \right|^2 \right] \\
	  &\qquad \le \RE(\Prob_{\A \ts} \| \Prob_{\A \tilde{\theta}}) + 2\,\pen(\tilde{\theta}) \\
	  &\qquad = \min_{\tb \in \Theta_L} \left[ \RE(\Prob_{\A \ts} \| \Prob_{\A \tb}) + 2\,\pen(\tb)  \right],
	\end{align*}
	which proves the lemma.
\end{IEEEproof}

\begin{lemma}
\label{l3}
If the estimators in $\Theta_L$ satisfy the condition \eqref{eq:strictly_positive}, then following inequality holds:
$$
 \RE(\Prob_{\A\ts}\parallel
      \Prob_{\A\tb}) \leq \frac{1}{c}\|\ts-\tb\|_1^2, \qquad \forall \tb \in \Theta_L.
$$

\end{lemma}
\begin{IEEEproof}
By definition of the KL divergence,
	\label{app:l3}
	\begin{align*}
	& \RE(\Prob_{\A \ts} \| \Prob_{\A \tb} ) \\
	& = \Exp_{\A \ts} \left[ \log \left( \frac{\Prob_{\A \ts}(y)}{\Prob_{\A \tb}(y)}\right)\right] \\
	& = \sum^m_{j=1} \Exp_{(\A \ts)_j} \left[ y_j \log \left( \frac{(\A \ts)_j}{(\A \tb)_j}\right)\right] \\
	& \qquad \qquad - \sum^m_{j=1} \Exp_{(\A \ts)_j} \left[ (\A \ts)_j - (\A \tb)_j\right] \\
	&= \sum^m_{j=1} \left[ (\A \ts)_j \log \left( \frac{(\A \ts)_j}{(\A \tb)_j}\right) - (\A \ts)_j + (\A \tb)_j\right] \\
	&\le \sum^m_{j=1} (\A \ts)_j \left(\frac{(\A \ts)_j}{(\A \tb)_j} -1 \right) - (\A \ts)_j + (\A \tb)_j \\
	&= \sum^m_{j=1} \frac{1}{(\A \tb)_j} \left| \left( \A \ts - \A \tb\right)_j \right|^2 \\
	&\le \frac{1}{c} \| \A(\ts - \tb) \|^2_2 \\
	&\le \frac{1}{c} \| \A(\ts - \tb) \|^2_1 \le \frac{1}{c} \| \ts - \tb \|^2_1.
	\end{align*}
	The first inequality uses $\log t\leq t-1$, the second is by \eqref{eq:strictly_positive}, the third uses the fact that the $\ell_1$ norm dominates the $\ell_2$ norm, and the last one is by the RIP-1 property (Lemma~\ref{RIP}).
\end{IEEEproof}

\begin{lemma}
\label{int}
Given two Poisson parameter vectors $\la,\lb \in \mathbb{R}^m_+$, the
following equality holds:
\begin{align*}
2 \log \frac{1}{\int\sqrt{\Prob_\la(\yb) \Prob_\lb(\yb)}d\mu(\yb)} = \sum_{j=1}^m
  \left|\la_j^{\half}-\lb_j^{\half}\right|^2, 
\end{align*}
where $\mu$ denotes the counting measure on $\Rplus^m$.
\end{lemma}
\begin{IEEEproof}
\begin{align*}
& \int \sqrt{\Prob_\la(\yb) \Prob_\lb(\yb)} d\mu(\yb) \\
& = \prod^m_{j=1} \sum^\infty_{y_j = 0} \frac{(\la_j \lb_j)^{y_j/2}}{\yb_j !} e^{-(\la_j + \lb_j)/2} \\
& = \prod^m_{j=1} e^{-\frac{1}{2}(\la_j - 2(\la_j \lb_j)^{1/2} + \lb_j)} \sum^\infty_{y_j = 0} \frac{(\la_j \lb_j)^{y_j/2}}{y_j !} e^{-(\la_j \lb_j)^{1/2}} \\
& = \prod^m_{j=1} e^{-\frac{1}{2}(\la_j - 2(\la_j \lb_j)^{1/2} + \lb_j)} \underbrace{\int \Prob_{(\la_j \lb_j)^{1/2}}(\yb_j) d\nu_j(y_j)}_{=1} \\
& = \prod^m_{j=1} e^{-\frac{1}{2}\left( \la_j^{1/2} - \lb_j^{1/2} \right)^2}
\end{align*}
Taking logs, we obtain the lemma.
\end{IEEEproof}

\section*{Acknowledgment}

\noindent The authors would like to thank Piotr
Indyk for his insightful comments on the performance of the expander
graphs, and the anonymous referees whose constructive criticism and numerous suggestions helped improve the quality of the paper.

\bibliographystyle{IEEEtran}
\bibliography{exp_pcs_R1.bbl}

\begin{IEEEbiographynophoto}{Maxim Raginsky} received the B.S.\ and M.S.\ degrees in 2000 and the
Ph.D.\ degree in 2002 from Northwestern University, Evanston, IL, all in electrical engineering. From 2002 to 2004 he was a Postdoctoral
Researcher at the Center for Photonic Communication and Computing at
Northwestern University, where he pursued work on quantum cryptography
and quantum communication and information theory. From 2004 to 2007 he
was a Beckman Foundation Postdoctoral Fellow at the University of
Illinois in Urbana-Champaign, where he carried out research on
information theory, statistical learning and computational
neuroscience. Since September 2007 he has been with Duke University, where he is now Assistant Research Professor of Electrical and Computer Engineering. His interests include statistical signal processing,
information theory, statistical learning and nonparametric
estimation. He is particularly interested in problems that combine the
communication, signal processing and machine learning components in a
novel and nontrivial way, as well as in the theory and practice
of robust statistical inference with limited information.
\end{IEEEbiographynophoto}

\begin{IEEEbiographynophoto}{Sina Jafarpour} is a Ph.D.\ candidate in the Computer Science Department of Princeton University, co-advised by Prof.\ Robert Calderbank and Prof.\ Robert Schapire. He received his B.Sc.\ in Computer Engineering from Sharif University of Technology in 2007. His main research interests include compressed sensing and applications of machine learning in image processing, multimedia, and information retrieval. Mr Jafarpour has been a member of the Van Gogh project supervised by Prof. Ingrid Daubechies since Fall 2008.
\end{IEEEbiographynophoto}

\begin{IEEEbiographynophoto}{Zachary Harmany} received the B.S.\ (magna cum lade) in Electrical Engineering and B.S.\ (cum lade) in Physics in 2006 from The Pennsylvania State University. Currently, he is a graduate student in the department of Electrical and Computer Engineering at Duke University. In 2010 he was a visiting researcher at The University of California, Merced. His research interests include nonlinear optimization, statistical signal processing, learning theory, and image processing with applications in functional neuroimaging, medical imaging, astronomy, and night vision. He is a student member of the IEEE as well as a member of SIAM and SPIE.
\end{IEEEbiographynophoto}

\begin{IEEEbiographynophoto}{Roummel Marcia} received his B.A.\ in Mathematics from Columbia  University in 1995 and his Ph.D.\ in
Mathematics from the University of California, San Diego in 2002.  He was a Computation and Informatics in Biology and Medicine Postdoctoral Fellow in the Biochemistry Department at the University of Wisconsin-Madison and a Research Scientist in the Electrical and Computer Engineering at Duke University.  He is currently an Assistant Professor of Applied Mathematics at the University of California, Merced.  His research interests include nonlinear optimization, numerical linear algebra, signal and image processing, and computational biology.
\end{IEEEbiographynophoto}

\begin{IEEEbiographynophoto}{Rebecca Willett} is an assistant professor in the Electrical and
Computer Engineering Department at Duke University. She completed her
Ph.D.\ in Electrical and Computer Engineering at Rice University in 2005.
Prof.\ Willett received the National Science Foundation CAREER Award in
2007, is a member of the DARPA Computer Science Study Group, and
received an Air Force Office of Scientific Research Young Investigator
Program award in 2010. Prof. Willett has also held visiting researcher
positions at the Institute for Pure and Applied Mathematics at UCLA in
2004, the University of Wisconsin-Madison 2003-2005, the French
National Institute for Research in Computer Science and Control
(INRIA) in 2003, and the Applied Science Research and Development
Laboratory at GE Healthcare in 2002. Her research interests include
network and imaging science with applications in medical imaging,
wireless sensor networks, astronomy, and social networks. Additional
information, including publications and software, are available online
at http://www.ee.duke.edu/~willett/.
\end{IEEEbiographynophoto}

\begin{IEEEbiographynophoto}{Robert Calderbank} received the B.Sc.\ degree in 1975 from Warwick University, England, the M.Sc.\ degree in 1976 from Oxford University, England, and the Ph.D.\ degree in 1980 from the California Institute of Technology, all in mathematics.

Dr.\ Calderbank is Dean of Natural Sciences at Duke University. He was previously Professor of Electrical Engineering and Mathematics at Princeton University where he directed the Program in Applied and Computational Mathematics. Prior to joining Princeton in 2004, he was Vice President for Research at AT\&T, responsible for directing the first industrial research lab in the world where the primary focus is data at scale.  At the start of his career at Bell Labs, innovations by Dr.\ Calderbank were incorporated in a progression of voiceband modem standards that moved communications practice close to the Shannon limit. Together with Peter Shor and colleagues at AT\&T Labs he showed that good quantum error correcting codes exist and developed the group theoretic framework for quantum error correction. He is a co-inventor of space-time codes for wireless communication, where correlation of signals across different transmit antennas is the key to reliable transmission.
 
Dr.\ Calderbank served as Editor in Chief of the IEEE Transactions on Information Theory from 1995 to 1998, and as Associate Editor for Coding Techniques from 1986 to 1989. He was a member of the Board of Governors of the IEEE Information Theory Society from 1991 to 1996 and from 2006 to 2008. Dr.\ Calderbank was honored by the IEEE Information Theory Prize Paper Award in 1995 for his work on the $Z_4$ linearity of Kerdock and Preparata Codes (joint with A.R.\ Hammons Jr., P.V.\ Kumar, N.J.A.\ Sloane, and P.\ Sole), and again in 1999 for the invention of space-time codes (joint with V.\ Tarokh and N.\ Seshadri). He received the 2006 IEEE Donald G.\ Fink Prize Paper Award and the IEEE Millennium Medal, and was elected to the U.S.\ National Academy of Engineering in 2005.
\end{IEEEbiographynophoto}

\end{document}